\documentclass[12pt]{article}

\pdfoutput=1

\usepackage[top=80pt,bottom=85pt,left=67pt,right=67pt]{geometry}
\usepackage{amssymb,amsmath,xypic}
\usepackage{graphicx,color,subfigure}
\usepackage{float}
\usepackage{sectsty}
\usepackage{cite}
\usepackage[debug,pageanchor=false]{hyperref}
\definecolor{link}{rgb}{.8,.15,.1}
\hypersetup{colorlinks=true,linkcolor=link,citecolor=link,urlcolor=link,linktocpage}

\makeatletter
\@addtoreset{equation}{section}
\makeatother

\renewcommand{\theequation}{\thesection.\arabic{equation}}

\newcommand{\beq}{\begin{equation}}
\newcommand{\eeq}{\end{equation}}
\newcommand{\nn}{\nonumber}
\newcommand{\Tr}{\mbox{Tr}}

\begin{document}

\begin{titlepage}

\begin{center}

\vskip .5in 
\noindent

{\Large \bf{Two dimensional ${\cal N}=(0,4)$  quivers dual to AdS$_3$ solutions in massive IIA   } }
\bigskip\medskip

Yolanda Lozano$^{a,}$\footnote{ylozano@uniovi.es},  Niall T. Macpherson$^{b,c,}$\footnote{ntmacpher@gmail.com}, Carlos Nunez$^{d,}$\footnote{c.nunez@swansea.ac.uk}, Anayeli Ramirez$^{a,}$\footnote{anayelam@gmail.com} \\

\bigskip\medskip
{\small

 $a$: Department of Physics, University of Oviedo,
Avda. Federico Garcia Lorca s/n, 33007 Oviedo, Spain
\vskip 3mm
 $b$: SISSA International School for Advanced Studies,
Via Bonomea 265, 34136 Trieste \\
		and \\INFN sezione di Trieste
		\vskip 3mm
$c$: International Institute of Physics, Universidade Federal do Rio Grande do Norte,
Campus Universitario - Lagoa Nova, Natal, RN, 59078-970, Brazil		
\vskip 3mm
 $d$: Department of Physics, Swansea University, Swansea SA2 8PP, United Kingdom}

\vskip .5cm 
\vskip .9cm 
     	{\bf Abstract }

\vskip .1in
\end{center}

\noindent
In this paper we discuss an infinite family of new solutions in massive Type IIA supergravity with AdS$_3\times$S$^2$ factors, preserving ${\cal N}=(0,4)$ SUSY. After studying  geometrical aspects of the backgrounds we propose a duality with a precise family of  quivers that  flow to (0,4) fixed points at low energies. These quivers consist on two families of (4,4) linear quivers coupled by matter fields. We present various tests of our proposal.
\noindent
\vskip .5cm
\vskip .5cm
\it{Dedicated to the memory of Steven S. Gubser. }
\vfill
\eject

\end{titlepage}

\setcounter{footnote}{0}

\tableofcontents
\newpage
\renewcommand{\theequation}{{\rm\thesection.\arabic{equation}}}
 \section{Introduction}
 The study of generic  quantum field theories (QFTs) is one of the main topics of interest in present-day  theoretical Physics. Perturbative and non-perturbative investigations in the recent decades have shown that remarkable progress can be achieved when the system under study is symmetric enough.
 
 One major line of research that came as a by-product of the Maldacena conjecture \cite{Maldacena:1997re}, is the study of supersymmetric and conformal field theories in diverse dimensions.
 Superconformal Field Theories (SCFTs) exist in space-time dimensions $d<7$ \cite{Nahm:1977tg}.
The last two decades witnessed  a large effort  in the classification of Type II or M-theory backgrounds with AdS$_{d+1}$ factors, see for example \cite{Gauntlett:2004zh},\cite{Gutowski:2014ova}. The  solutions are conjectured to be dual to SCFTs in $d$ dimensions with different amounts of SUSY. In the case in which we have eight Poincar\'e supercharges major progress has been achieved (the number of real supercharges doubles by the presence of the conformal partner supercharges).

For the case of ${\cal N}=2$ SCFTs in four dimensions, the field theories studied in \cite{Gaiotto:2009we} have holographic duals first discussed in \cite{Gaiotto:2009gz}, and further elaborated (among other works) in \cite{ReidEdwards:2010qs}-\cite{Bah:2019jts}. The case of five dimensional SCFTs was analysed from the field theoretical and holographic viewpoints in \cite{DHoker:2016ujz}-\cite{Bergman:2018hin}, among many other interesting works.   An infinite family of six-dimensional ${\cal N}=(1,0)$ SCFTs was discussed from both the field theoretical and holographic points of view in \cite{Apruzzi:2015wna}-\cite{Hanany:1997gh}. For three-dimensional  ${\cal N}=4$ SCFTs, the field theories presented in \cite{Gaiotto:2008ak} were discussed holographically in \cite{DHoker:2007hhe}-\cite{Lozano:2016wrs}, among other works.

The case of  two-dimensional SCFTs  and their AdS duals  is particularly attractive.  The interest that CFTs in two dimensions and  AdS$_3$ solutions present in other areas of theoretical Physics (condensed matter systems, black holes, etc),  and  the power of the 2-d super conformal algebra present us with a perfect {\it theoretical lab} to test various ideas explicitly. This motivated various
 attempts at finding classifications of AdS$_3$  backgrounds and studying their dual CFTs --for a sample of papers see
 \cite{Witten:1997yu}-\cite{Lozano:2019emq}.

In this work we add a new entry to the dictionary between SCFTs and string backgrounds with an AdS-factor described above. We deal with ${\cal N}=(0,4)$ (small algebra) SCFTs. We define our SCFTs as the IR fixed points of ${\cal N}=(0,4)$ UV finite QFTs. These QFTs
 are
described by quivers, consisting of two long rows of gauge groups connected by hypermultiplets and Fermi multiplets. There are also global (flavour) symmetry groups, joined with the gauge groups by Fermi multiplets. Quantum theories of this kind (with some differences regarding the field content and R-symmetry charges) have been proposed in the study of solitonic strings in six-dimensional ${\cal N}=(1,0)$ SCFTs, see for example \cite{Gadde:2015tra}\footnote{See also \cite{Hanany:2018hlz} for realisations in terms of D3-brane boxes.}.
We show that the new background solutions to massive IIA supergravity constructed recently in \cite{Lozano:2019emq} contain the needed isometries to be dual to our SCFTs. These backgrounds may be trusted when the number of nodes of the quiver is large and so are the ranks of each gauge group\footnote{See the recent paper  \cite{Uhlemann:2019ypp} for long 5d quivers.}. We show that they reproduce the central charge of our SCFTs in the holographic limit.

The contents of this paper are distributed as follows. In section \ref{geometria} we summarise the general massive Type IIA backgrounds that we constructed recently in \cite{Lozano:2019emq}, and find new solutions, also presented in \cite{Lozano:2019jza}. These backgrounds have the structure
\begin{equation}
\text{AdS}_3\times \text{CY}_2\times\text{S}^2\times \text{I}_\rho.
\end{equation}
By I$_\rho$ we denote an interval parametrised by a coordinate that we label $\rho$. There are warp factors in front of each metric component (also for each of the RR and NS fluxes compatible with the isometries of the background). 
We discuss various observable quantities of these backgrounds, like the Page charges, the explicit presence of branes (we map these data into  Hanany-Witten brane set-ups) and the holographic central charges. All these quantities are described in terms of the functions that define the warp factors.
\\
In section \ref{CFTsect} we define the QFTs of our interest. In order to do this we take a small detour through 2-d ${\cal N}=(0,2)$ 
multiplets. In terms of them we write the field content of our ${\cal N}=(0,4)$ QFTs. We pay special attention to the cancellation of gauge anomalies. We propose that these QFTs flow in the IR to strongly coupled ${\cal N}=(0,4) $ SCFTs with small superconformal algebra. We use this to link the R-symmetry anomaly (the level of the Kac-Moody algebra) with the central charge (the leading coefficient in the OPE of energy-momentum tensors). We finally propose a generic duality between our SCFTs 
 and the backgrounds discussed in section \ref{geometria}.
 \\
 In section \ref{ejempl} (of pedagogical character), we present a detailed  set of examples that serve as tests of our proposed duality. In those examples we show how the supergravity backgrounds (with the predicted number of {\it colour} and {\it flavour} branes) have the precise {\it combinatorics} to be dual to long quivers with non-anomalous gauge symmetries and flavour symmetries. We calculate the central charge in the SCFT and the holographic central charge in  the gravity background
showing a clean matching between both descriptions. 

We close the paper with a brief summary and some ideas for further research in section \ref{conclusion}. The presentation is complemented by appendixes of technical nature.

\section{The holographic backgrounds}\label{geometria}
In this section we start by discussing  the solutions to massive IIA supergravity (with localised sources) obtained in the recent work \cite{Lozano:2019emq}. We propose that these backgrounds are
holographic duals to two dimensional CFTs preserving ${\cal N}=(0,4)$ SUSY. The particular CFTs will be discussed in section \ref{CFTsect}. The Neveu-Schwarz (NS) sector of these bosonic solutions reads,
\begin{align}
ds^2&= \frac{u}{\sqrt{\hat{h}_4 h_8}}\bigg(ds^2(\text{AdS}_3)+\frac{h_8\hat{h}_4 }{4 h_8\hat{h}_4+(u')^2}ds^2(\text{S}^2)\bigg)+ \sqrt{\frac{\hat{h}_4}{h_8}}ds^2(\text{CY}_2)+ \frac{\sqrt{\hat{h}_4 h_8}}{u} d\rho^2,\\
e^{-\Phi}&= \frac{h_8^{\frac{3}{4}} }{2\hat{h}_4^{\frac{1}{4}}\sqrt{u}}\sqrt{4h_8 \hat{h}_4+(u')^2},~~~~ H= \frac{1}{2}d(-\rho+\frac{ u u'}{4 \hat{h}_4 h_8+ (u')^2})\wedge\text{vol}(\text{S}^2)+ \frac{1}{h_8}d\rho\wedge H_2,\nn
\end{align}
here $\Phi$ is the dilaton, $H=dB_2$ is the NS 3-form and $ds^2$ is written in string frame. The warping function $\hat{h}_4$ has support on $(\rho,\text{CY}_2)$. On the other hand,  $u$ and $h_8$ only depend of $\rho$. We denote $u'= \partial_{\rho}u$ and similarly for $h_8'$. 
The RR fluxes are 
\begin{subequations}
\begin{align}
F_0&=h_8',\;\;\;F_2=-H_2-\frac{1}{2}\bigg(h_8- \frac{ h'_8 u'u}{4 h_8 \hat{h}_4+ (u')^2} \bigg)\text{vol}(\text{S}^2),\label{eq:classIflux2}\\[2mm]
F_4&= \bigg(d\left(\frac{u u'}{2\hat{ h}_4}\right)+2 h_8  d\rho\bigg) \wedge\text{vol}(\text{AdS}_3)\nn\\[2mm]
& -\frac{h_8}{u} (\hat \star_4 d_4 \hat{h}_4)\wedge d\rho- \partial_{\rho}\hat{h}_4\text{vol}(\text{CY}_2)-\frac{u u'}{2 ( 4h_8 \hat{h}_4+ (u')^2)} H_2\wedge \text{vol}(\text{S}^2),\label{eq:classIflux3}
\end{align}
\end{subequations}
with the higher fluxes related to them as $F_6=-\star_{10} F_4,~F_8=\star_{10} F_2,~F_{10}=-\star_{10} F_0$.
It was shown in \cite{Lozano:2019emq} that supersymmetry holds whenever
\beq
u''=0,~~~~ H_2+ \hat{\star}_4 H_2=0,\label{messi}
\eeq
where $\hat{\star}_4$ is the Hodge dual on CY$_2$. In what follows we will concentrate on the set of solutions for which $H_2=0$.
The Bianchi identities of the fluxes then impose (away from localised sources)
\begin{align}\label{eq:BI1}
&h''_8= 0,\;\;\;\;\frac{h_8}{u}\nabla^2_{\text{CY}_2}\hat{h}_4+ \partial_{\rho}^2 \hat{h}_4 =0.\nn
\end{align}
A further restriction consists in assuming that $\hat{h}_4=\hat{h}_4(\rho)$. 
After this, the string frame background reads,
\begin{eqnarray}
ds_{st}^2&=& \frac{u}{\sqrt{\hat{h}_4 h_8}}\bigg(ds^2(\text{AdS}_3)+\frac{h_8\hat{h}_4 }{4 h_8 \hat{h}_4+(u')^2}ds^2(\text{S}^2)\bigg)+ \sqrt{\frac{\hat{h}_4}{h_8}}ds^2(\text{CY}_2)+ \frac{\sqrt{\hat{h}_4 h_8}}{u} d\rho^2,\nn \\
e^{-\Phi}&=& \frac{h_8^{\frac{3}{4}} }{2\hat{h}_4^{\frac{1}{4}}\sqrt{u}}\sqrt{4h_8 \hat{h}_4+(u')^2},~~~~~~ B_2= \frac{1}{2}\left(-\rho+ 2\pi k+\frac{ u u'}{4 \hat{h}_4 h_8+ (u')^2} \right) \text{vol}(\text{S}^2),\nn\\
\hat{F}_0&=&h_8',\;\;\;\;
\hat{F}_2=-\frac{1}{2}\bigg(h_8- h_8'(\rho-2\pi k)\bigg)\text{vol}(\text{S}^2),\nn\\
\hat{F}_4&=& \bigg(\partial_\rho\left(\frac{u u'}{2 \hat{h}_4}\right)+2 h_8\bigg)  d\rho \wedge\text{vol}(\text{AdS}_3)
- \partial_{\rho}\hat{h}_4\text{vol}(\text{CY}_2).\label{eq:background}
\end{eqnarray}
We have written the Page fluxes $\hat{F}=e^{-B_2}\wedge F$ that are more useful for our purposes. Notice that we have also allowed for large gauge transformations $B_2\to B_2 + { \pi k} \text{vol}(\text{S}^2)$, for $k=0,1,...., P$. The transformations are performed every time we cross an interval $[2\pi k, 2\pi(k+1)]$. To motivate this consider the following: in the limit where $\hat{h}_4(\rho)$ and/or $h_8(\rho)$ become large compared with $u(\rho)$ the NS 2-form in the presence of $k$ large gauge transformations is approximately
\beq
B_2\sim \frac{1}{2}(-\rho+ 2\pi k)\text{vol}(\text{S}^2) \implies \hat{b}_0 = -\frac{1}{(2\pi)^2}\int_{\text{S}^2}B_2\sim \frac{1}{2\pi}(\rho-2\pi k).
\eeq
This can be archived by tuning certain integration constants in the solutions presented below, and in fact coincides  with the limit of weak curvature where the supergravity approximation can be trusted. Demanding that $\hat{b}_0$ lies in the fundamental region $\hat{b}_0\in[0,1)$ partitions the real line spanned by $\rho$ into segments of length $2\pi$. A  large gauge transformation ( $B_2\to B_2+ \pi \text{vol}(\text{S}^2)$) is required as one crosses between these segments, such that the NS 2-form quoted in \eqref{eq:background} is valid in the segment $2k\pi\leq \rho <2\pi(k+1)$ with $k=0,1,2...$.

The background in \eqref{eq:background} is a SUSY solution  of the massive IIA equations of motion if the functions $\hat{h}_4,h_8,u$ satisfy (away from localised sources),
\begin{equation}
\hat{h}_4''(\rho)=0,\;\;\;\; h_8''(\rho)=0,\;\;\;\; u''(\rho)=0.\label{eqsmotion}
\end{equation}
The three functions are thus linear. Various particular solutions were analysed in \cite{Lozano:2019emq}. Here we will  present an infinite family of solutions for which the functions are piecewise continuous.  
%
\subsection{The local solutions}\label{soluciones}
We shall be interested in solutions that in the interval $2\pi k\leq \rho\leq 2\pi (k+1)$ (for $k=0,1,....,P$) are of the form,
\begin{equation}
\hat{h}_4^{(k)}=\Upsilon \left( \alpha_k +\frac{\beta_k}{2\pi}(\rho-2\pi k)\right),\;\;\;\; h_8^{(k)}= \mu_k +\frac{\nu_k}{2\pi}(\rho-2\pi k),\;\;\;u^{(k)}= a_k +\frac{b_k}{2\pi}(\rho-2\pi k).\nonumber
\end{equation}
Here ($\Upsilon, \alpha_k,\beta_k, \mu_k,\nu_k, a_k,b_k$) are arbitrary constants whose physical meaning we shall discuss below.
In particular, we impose that these three functions vanish at $\rho=0$ (where the space begins) and that the space ends at $\rho=2\pi(P+1)$, by considering the situation for which $\hat{h}_4$ and/or $h_8$ vanish at this point. These conditions leave us with functions of the form,

  \begin{equation} \label{profileh4final}
\hat{h}_4(\rho)=\Upsilon h_4(\rho)
                    =\Upsilon \left\{ \begin{array}{ccrcl}
                       \frac{\beta_0 }{2\pi}
                       \rho & 0\leq \rho\leq 2\pi \\
                                      \alpha_k + \frac{\beta_k}{2\pi}(\rho-2\pi k ) &~~ 2\pi k\leq \rho \leq 2\pi(k+1),\;\;\;\; k:=1,....,P-1\\
                      \alpha_P+  \frac{\beta_P}{2\pi}(\rho-2\pi P) & 2\pi P\leq \rho \leq 2\pi(P+1).
                                             \end{array}
\right.
\end{equation}
 \begin{equation} \label{profileh8final}
h_8(\rho)
                    =\left\{ \begin{array}{ccrcl}
                       \frac{\nu_0 }{2\pi}
                       \rho & 0\leq \rho\leq 2\pi \\
                                      \mu_k + \frac{\nu_k}{2\pi}(\rho-2\pi k) &~~ 2\pi k\leq \rho \leq 2\pi(k+1),\;\;\;\; k:=1,....,P-1\\
                      \mu_P+  \frac{\nu_P}{2\pi}(\rho-2\pi P) & 2\pi P\leq \rho \leq 2\pi(P+1).
                                             \end{array}
\right.
\end{equation}
  \begin{equation} \label{profileufinal}
u(\rho)
                    =\left\{ \begin{array}{ccrcl}
                       \frac{b_0 }{2\pi}
                       \rho & 0\leq \rho\leq 2\pi \\
                                      a_k + \frac{b_k}{2\pi}(\rho-2\pi k) &~~ 2\pi k\leq \rho \leq 2\pi(k+1),\;\;\;\; k:=1,....,P-1\\
                      a_P+ \frac{b_P}{2\pi}(\rho-2\pi P) & 2\pi P\leq \rho \leq 2\pi(P+1).
                                             \end{array}
\right.
\end{equation}
If the function $\hat{h}_4(\rho)$ vanishes at $\rho=2\pi(P+1)$, ending the space there, we need that $\alpha_P=-\beta_P$. Similarly if $h_8\big(2\pi(P+1)\big)=0$, we must impose that $\nu_P=-\mu_P$. 
\\
Demanding that the metric, dilaton and $B_2$ field are continuous across the different intervals imposes additional conditions on the various constants\footnote{We do not impose the continuity of $H= dB_2$ since $H=F(\rho) d\rho \wedge  \text{vol}(\text{S}^2)$. This implies that $dH=0$  and the continuity of $H$ is not needed to avoid the presence of NS brane sources.  }. The details are discussed in appendix \ref{continuityconditions}. Here we quote one simple solution to these continuity equations,
\begin{equation}
\mu_k=\sum_{j=0}^{k-1} \nu_j,\;\;\; \alpha_k=\sum_{j=0}^{k-1}\beta_j,\;\;\;\; b_k=b_0,\;\;\;\; a_k= k b_0.\label{conticond}
\end{equation}
These conditions imply the continuity of  the functions $\hat{h}_4, h_8$. Their derivatives can, however, present jumps. This will imply discontinuities in the RR sector, that we will interpret  as generated by the presence of branes in the background, that modify the Bianchi identities.  In turn, notice that \eqref{conticond} implies that $u(\rho)=\frac{ b_0}{2\pi} \rho$ in all intervals,
which is consistent with the supersymmetry requirement \eqref{messi} that $u''=0$ globally.
\\
  These supergravity backgrounds can be trusted (with localised singularities) if the numbers $P, \alpha_k,\mu_k$ are large. Indeed, the Ricci scalar only diverges at the points where the sources are localised. Choosing the numbers $\nu_k, \beta_k$ to be large controls this divergence. On the other hand $P$ is taken to be large to have these
singularities separated enough that we can trust the geometric description given here.
\subsection{The $\rho$-interval}
Let us analyse more closely these solutions. The background functions defined in the first interval $[0, 2\pi]$ show that the space begins at $\rho=0$ in a smooth fashion.
On the other hand, the $\rho$-interval 
ends at a generic point $\rho= 2\pi(P+1)$ if  any of the functions $\hat{h}_4$ and/or $h_8$  vanish at that point. Let us analyse the behaviour of the metric and dilaton close to the end of the space for the three possible cases:
\begin{itemize}
\item The space ends by virtue of the function $\hat{h}_4$ whilst $h_8$ is generically non-vanishing at $\rho=2\pi(P+1)$. In the last interval the functions defining the background are  then
\begin{equation}
\hat{h}_4=\Upsilon\left( \alpha_P-\frac{\alpha_P}{2\pi} (\rho-2\pi P)\right),\;\;\;h_8= \mu_P+\frac{\nu_P}{2\pi} (\rho-2\pi P),\;\;\;u=\frac{b_0}{2\pi} \rho.\nonumber
\end{equation} 
In this case, expanding the metric and the dilaton close to $\rho=2\pi(P+1)$ we find, for small values of $x=2\pi(P+1) -\rho$,
\beq
ds^2\sim \frac{m_1}{\sqrt{x}} ds^2(\text{AdS}_3) +\frac{\sqrt{x}}{m_1}\Big[dx^2+ m_1 m_2 ds^2(\text{S}^2) + m_3 m_1 ds^2(\text{CY}_2)  \Big],\;\;\; e^{-4\Phi}=\frac{m_4}{x}.\label{cani}
\eeq
The numbers $(m_1,....,m_4)$ are written in terms of $\mu_P,\alpha_P,\nu_P, b_0,\Upsilon$. This asymptotic behaviour indicates that close to the end of the space we have a D2 brane that extends on AdS$_3$ and is delocalised (or smeared) on CY$_2 \times $S$^2$---see \cite{Lozano:2019emq} for a generic analysis of singularities. Note that one could also view this as an O2 plane smeared on CY$_2 \times $S$^2$ or a superposition of both D2s and O2s. \\
\item The space ends by virtue of the function $h_8$ while $\hat{h}_4$ is generically non-vanishing at $\rho=2\pi(P+1)$. In the last interval the functions are then
\begin{equation}
h_8=\mu_P-\frac{\mu_P}{2\pi}(\rho-2\pi P),\;\;\;\; \hat{h}_4=\Upsilon\left(\alpha_P+\frac{\beta_P }{2\pi}(\rho-2\pi P)\right),\;\;\;\; u=\frac{b_0}{2\pi} \rho.\nonumber
\end{equation}
For small $x=2\pi(P+1)-\rho$, the metric and dilaton scale as,
\beq
 ds^2\sim \frac{1}{\sqrt{x}}\Big[n_1 ds^2(\text{AdS}_3)+ n_3 ds^2({\text{CY}_2})    \Big] +\frac{\sqrt{x}}{n_1}\Big[dx^2+ n_1 n_2 ds^2(\text{S}^2)  \Big],\;\;\;e^{-4\Phi}=n_4 x^3.\label{diego}
\eeq
The numbers $(n_1,....,n_4)$ are written in terms of $\mu_P,\alpha_P,\beta_P, b_0,\Upsilon$. This asymptotic behaviour indicates that at $\rho=2\pi(P+1)$ we have an O6 plane that extends on AdS$_3\times$CY$_2$.\\
\item Finally, consider the more symmetric  case for which the space  is closed by the simultaneous vanishing of $\hat{h}_4$ and $ h_8$ at $\rho=2\pi(P+1)$. In this case the functions in the last interval read,
\begin{equation}
h_8=\mu_P-\frac{\mu_P}{2\pi}(\rho-2\pi P),\;\;\;\; \hat{h}_4=\Upsilon\left(\alpha_P-\frac{\alpha_P}{2\pi} (\rho-2\pi P)\right),\;\;\; u=\frac{b_0}{2\pi} \rho.\label{bati}
\end{equation}
For small values of  $x=2\pi(P+1)-\rho$, the metric and dilaton scale as,
\beq
 ds^2\sim \frac{s_1}{{x}} ds^2(\text{AdS}_3)+ s_3 ds^2(\text{CY}_2)   +\frac{ {x}}{s_1}\Big[dx^2+ s_1 s_2 ds^2(\text{S}^2)
   \Big],\;\;
 e^{-4\Phi}=s_4 x^2.
\eeq
The numbers $(s_1,....,s_4)$ are written in terms of $\mu_P,\alpha_P, b_0,\Upsilon$. Notice that each quantity above is the product of those in \eqref{cani}-\eqref{diego}. This indicates the superposition of O2-O6 planes.\\ 
~\\
This more symmetric way of ending the space is the one 
on which we will concentrate our forthcoming analysis. An important observation is that, from the gravity perspective, the behaviour we are finding close to the end of the interval is the least healthy of the three analysed, as the O2s need to be smeared. We believe that the presence of smeared O-planes is an artifact of the supergravity approximation.\\
~\\
To be used below, let us quote the explicit expressions for the different numerical values of $(s_1,s_2,s_3,s_4)$,
\begin{eqnarray}
& & s_1= \frac{4\pi^2 b_0 (P+1)}{\sqrt{\alpha_P\mu_P \Upsilon}},\;\;\;\; s_2=2\pi(P+1)\frac{\sqrt{\alpha_P\mu_P \Upsilon}}{b_0},\nn\\
& & s_3= \sqrt{\frac{\Upsilon \alpha_P}{\mu_P}},\;\;\;\; s_4=\frac{b_0^2 \mu_P^3}{2^{10}\pi^6 \alpha_P (P+1)^2 \Upsilon}.\label{lio}
\end{eqnarray}
Notice that in order for the CY$_2$ space to be large compared with the string size, we need that $\Upsilon \alpha_P\sim \mu_P$. Otherwise the gravity background is not trustable.\\
\end{itemize}

In the following section we study the Page charges and discuss the presence of branes in our solutions. These are of the form given by eq.\eqref{eq:background}, with the functions $({\hat h}_4, h_8, u)$ satisfying eq.\eqref{eqsmotion}, 
away from localised sources, and piecewise continuous, as in \eqref{profileh4final}-\eqref{profileufinal}. The condition for continuity
of the defining functions $\hat{h}_4, h_8$ is given by \eqref{conticond}. This implies the continuity of the NS-sector of the solution. From all the possibilities to end the space we focus on solutions whose last interval's functions are given by   \eqref{bati}. The non-compact solution with $\hat{h}_4\sim h_8\sim u\sim \rho$ all over the space will be discussed in detail in \cite{LMNR3}.

\subsection{Page charges}
The Page charges are important observable quantities characterising a supergravity solution. Since they are quantised they imply the quantisation of some of the constants defining the solution in \eqref{profileh4final}-\eqref{profileufinal}. The Page charge of Dp-branes is given by the integral of the magnetic part of the Page $\hat{F}_{8-p}$ form. This is,
\begin{equation}
{(2\pi)^{7-p} g_s \alpha'^{(7-p)/2}}Q_{Dp}= \int_{\Sigma_{8-p}} \hat{F}_{8-p}.\label{pagedef}
\end{equation}
In what follows, we choose units consistent with $\alpha'=g_s=1$. 
Also, we will use that $\hat{h}_4=\Upsilon h_4$, as seen in \eqref{profileh4final}.
\\
We find the following Page charges for our solutions in the interval $[2\pi k, 2\pi(k+1)]$,
\begin{eqnarray}
& & Q_{D8}=2\pi  F_0 =2\pi h_8'=  \nu_k.\label{cargasxx}\\
& & Q_{D6}=\frac{1}{2\pi}\int_{\text{S}^2} \hat{F}_2= {h_8- {h_8'}(\rho-2\pi k)}=\mu_k.\nonumber\\
& & Q_{D4}=\frac{1}{8\pi^3}\int_{\text{CY}_2} \hat{F}_4=\Upsilon \frac{\text{Vol(CY}_2)}{16\pi^4} \beta_k,\nonumber\\
& & Q_{D2}=\frac{1}{32\pi^5}\int_{\text{CY}_2\times \text{S}^2} \hat{F}_6=\Upsilon \frac{\text{Vol(CY}_2)}{16\pi^4} (h_4- h_4'(\rho-2\pi k))= \Upsilon \frac{\text{Vol(CY}_2)}{16\pi^4} \alpha_k .\nonumber
\end{eqnarray}
We have used that the magnetic part of $\hat{F}_6$ is 
\begin{equation}
\hat{F}_{6,mag}=\hat{f}_6=\frac{\Upsilon }{2}\left(h_4- h_4'(\rho-2\pi k)\right) \text{vol}(\text{S}^2)\wedge \text{vol}(\text{CY}_2) .\label{estax}
\end{equation} 
We also have one NS-five brane every time we cross the value $\rho=2\pi k$ (for $k=1,....,P$). The total number of NS-five branes is $Q_{NS}=\frac{1}{4\pi^2}\int_{\rho\times S^2}H_3=(P+1)$.\\
 In what follows, we  choose the constant $\Upsilon$ to satisfy $\Upsilon \text{Vol(CY}_2) =16\pi^4$.
This implies that the constants $\alpha_k,\beta_k$ are integer numbers (like $\nu_k,\mu_k$ are). They are directly related with the number of branes in the associated Hanany-Witten brane set-up.\\
~\\
To understand which branes are present in our backgrounds, let us study the Bianchi identities for the Page fluxes.
\subsubsection{Hanany-Witten brane set-up}
We now calculate the Bianchi identities for the Page fluxes. The goal is to determine which branes are actually present in our background solutions, either as sources or dissolved into
 fluxes.
 
\begin{figure}[h!]
    \centering

    {{\includegraphics[width=10cm]{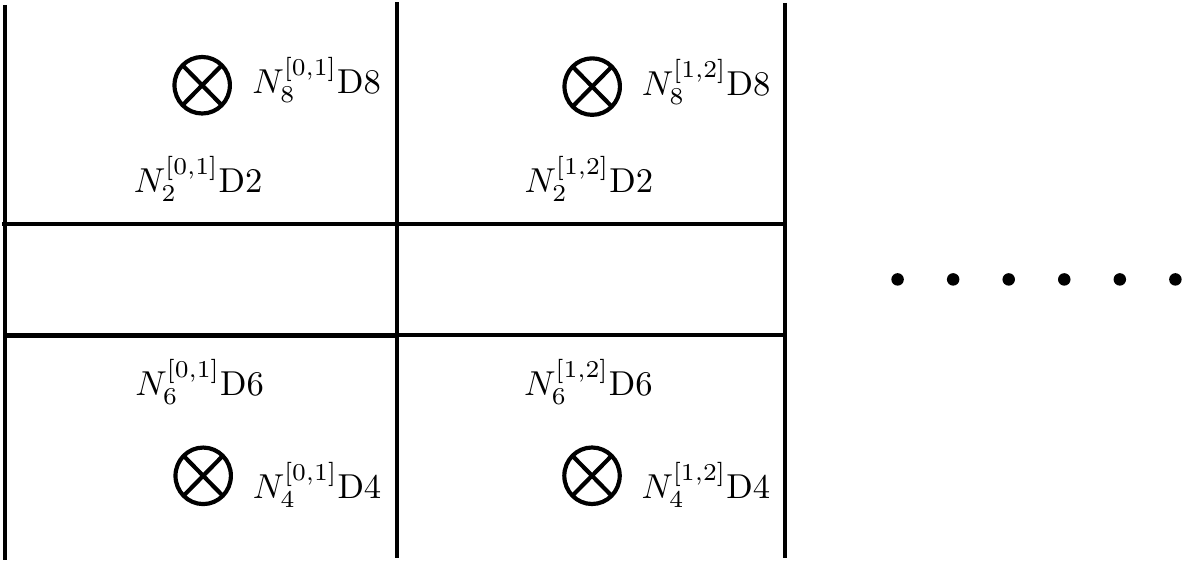} }}%

\caption{The generic Hanany-Witten set-up associated with our backgrounds. The vertical lines are NS-five branes. The horizontal lines represent D2 and D6 branes. The crosses indicate D4 and D8 branes.}
\label{xxy}

\end{figure}

Let us start with the flux $F_0=h_8'(\rho)$. We calculate
$dF_0= h_8''(\rho) d\rho$. Now, at a generic point of the $\rho$-coordinate we will have  $h_8''=0$, according to  \eqref{eqsmotion}. However, due to our  definition of the functions $\hat{h}_4$ and $h_8$---see \eqref{profileh4final}-\eqref{profileh8final}, something special occurs at the points where the functions change slope. In fact, for both $\hat{h}_4$ and $h_8$ we find,
\begin{equation}
h_8''=\sum_{k=1}^{P} \left(\frac{\nu_{k-1}-\nu_{k}}{2\pi}\right)\delta(\rho-2\pi k),\;\;\; \hat{h}_4''=\Upsilon \sum_{k=1}^{P} \left(\frac{\beta_{k-1}-\beta_{k}}{2\pi}\right)\delta(\rho-2\pi k).\label{carlitos}
\end{equation}
As a  consequence of this we have,
\begin{eqnarray}
& & dF_0=\sum_{k=1}^{P} \left(\frac{\nu_{k-1}-\nu_{k}}{2\pi}\right)\delta(\rho-2\pi k) d\rho,\label{secondd}\\
& & d\hat{F}_4= \Upsilon \sum_{k=1}^{P} \left(\frac{\beta_{k-1}-\beta_{k}}{2\pi}\right)\delta(\rho-2\pi k) d\rho\wedge \text{vol}(\text{CY}_2),\nn
\end{eqnarray}
indicating that at the points $\rho=2 \pi k$ there may be localised D8 and semi-localised D4 branes.
In fact, explicit D8 and D4 branes are present at $\rho=2\pi k$ when the slopes of $h_8,\hat{ h}_4$ are different at both sides.\\
Let us investigate the same about D2 and D6 branes. For the magnetic part of the Page fluxes, we compute in the interval $[2\pi k, 2\pi (k+1)]$
\begin{eqnarray}
& & d\hat{F}_2=\frac{1}{2} h_8''\times  (\rho-2\pi k) d\rho\wedge \text{vol}(\text{S}^2),\label{zada}\\
& & d\hat{F}_6= d\hat{f}_6= \frac{1}{2} {\hat{h}}_4'' \times(\rho-2\pi k) d\rho\wedge \text{vol}(\text{S}^2)\wedge \text{vol}(\text{CY}_2).\nn
\end{eqnarray}
Using  \eqref{carlitos} and that $x\delta(x)=0$, we then find that there are no 
sources for D2 or D6 branes present. 
This is precisely because a large gauge
transformation of the NS two-form is performed at the loci of the D8 and D4s, were this not
the case a source term for D6 and D2 would be induced as in section 5.1 of \cite{Lozano:2019emq} \footnote{The D8 and D4 can also be 
shown to be supersymmetric by a small modification of
the argument in  \cite{Lozano:2019emq}. There, it was assumed that no gauge transformations are performed
on the brane, which lead to D8 and D4 world volume gauge fields being required by
supersymmetry and the source corrected Bianchi identities. Here these gauge fields have
been absorbed by the large gauge transformation of the NS two-form.
The branes now restricted to lie at $\rho= 2\pi (k+1)$, $k=0,1,2...$. We give some details in Appendix \ref{apendice2}.}.
\\
This study suggests that the D2 and D6 branes will play the role of {\it colour} branes, while the D4 and D8 branes that of {\it flavour} branes. The global symmetry in the dual CFT is gravitationally realised by the gauge fields that fluctuate on the D4 or D8 branes.
\\
Studying the associated Hanany-Witten \cite{Hanany:1996ie} set-up, we find that in flat space the branes are distributed as
indicated in table \ref{D6-NS5-D8-D2-D4-first}.
 \begin{table}[ht]
	\begin{center}
		\begin{tabular}{| l | c | c | c | c| c | c| c | c| c | c |}
			\hline		    
			& 0 & 1 & 2 & 3 & 4 & 5 & 6 & 7 & 8 & 9 \\ \hline
			D2 & x & x & &  &  &  & x  &   &   &   \\ \hline
			D4 & x & x &  &  &  &   &  & x & x & x  \\ \hline
			D6 & x & x & x & x & x & x & x  &   &   &   \\ \hline
			D8 & x & x &x  & x & x &  x &  & x & x & x  \\ \hline
			NS5 & x & x &x  & x & x & x  &   &   &  &  \\ \hline
		\end{tabular} 
	\end{center}
	\caption{$\frac18$-BPS brane intersection underlying our geometry. The directions $(x^0,x^1)$ are the directions where the 2d CFT lives  (dual to our AdS$_3$). The directions $(x^2, \dots, x^5)$ span the CY$_2$, on which the D6 and the D8-branes are wrapped. The coordinate $x^6$ is the direction associated with $\rho$. Finally $(x^7,x^8,x^9)$ are the transverse directions realising an SO(3)-symmetry associated with the isometries of S$^2$.}   
	\label{D6-NS5-D8-D2-D4-first}	
\end{table} 
Our proposal is that  the geometries described by \eqref{eq:background}, capture the near horizon, or decoupling limit, of the brane configuration, once a suitable large number of NS and D-branes is considered.
\\
Using our result for the Page charges in \eqref{cargasxx} and the modified Bianchi identities in \eqref{secondd}, we find 
that the number of D-branes in the interval $[2\pi (k-1) , 2\pi k]$ (in between two NS-five branes) is,
\begin{eqnarray}
& & N_{D8}^{[k-1,k]}= \nu_{k-1}-\nu_k,\;\;\;\;\; N_{D4}^{[k-1,k]}= \beta_{k-1}-\beta_k,\label{numberofbranes}\\
& & 
N_{D6}^{[k-1,k]}= \mu_k=\sum_{i=0}^{k-1}\nu_i,\;\;\;\; N_{D2}^{[k-1,k]}= \alpha_k=\sum_{i=0}^{k-1}\beta_i.\label{numberofcolour}
\end{eqnarray}
We then have a Hanany-Witten brane set-up, that in the interval $[2\pi(k-1),2\pi k]$ (bounded by NS-five branes), has $N_{D6}^{[k-1,k]}, N_{D2}^{[k-1,k]}$ colour branes and  $N_{D8}^{[k-1,k]}, N_{D4}^{[k-1,k]}$ flavour branes.  See figure \ref{xxy}.

\subsection{Holographic central charge}
To close our  study of the background in \eqref{eq:background} we will calculate the holographic central charge associated with these solutions. The idea is to compare with the central charge of the proposed dual conformal field theory, that we study in the coming sections.
\\
The central charge is one of the important observables for conformal field theories. It appears when calculating the trace of the energy-momentum tensor, for a theory defined on a curved space. In the case of two dimensional conformal field theories, there is only one relevant quantity -- denoted by $"c"$--that appears when computing $<T_\mu^\mu>=-\frac{c}{24\pi} R$. Here $R$ is the Ricci scalar of the manifold on which the CFT is defined and $c$ is the central charge. \\
The holographic calculation of this quantity has a very interesting history. It was first obtained  in
\cite{Brown:1986nw} (before the Maldacena conjecture was formulated),
then calculated in \cite{Henningson:1998gx}. In the context of AdS-supergravity, it  was holographically computed in \cite{Freedman:1999gp} and \cite{Kraus:2005zm}. In  \cite{Klebanov:2007ws} generic supergravity solutions were considered that were later generalised in \cite{Macpherson:2014eza}. This is the formalism we will use. It basically boils down to computing  the volume of the
internal space (excluding AdS$_3$). 
\\
In a putative compactification to an effective 3-d supergravity this volume is the inverse of the 3-d Newton constant. However, in general, it needs to be weighted by factors of the dilaton and other warp factors. In fact, for a generic dilaton and  background of the form,
\begin{equation}
ds^2= a(r,\vec{\theta})(dx_{1,d}^2 + b(r)dr^2) + g_{ij}(r,\vec{\theta}) d\theta^id\theta^j,\;\;\;\; \Phi(r,\vec{\theta}),
\end{equation}
one should calculate the auxiliary quantity \cite{Macpherson:2014eza}
\begin{equation}
\hat{H}= \left(\int d\vec{\theta} \sqrt{e^{-4\Phi} \det[g_{ij}] a(r,\vec{\theta})^d } \right)^2.\nn
\end{equation}
With this, one computes the holographic central charge (see \cite{Macpherson:2014eza,Bea:2015fja} for the derivation) to be,
\begin{equation}
c_{hol}= 3\times \frac{ d^d}{ G_N} \frac{b(r)^{d/2} (\hat{H})^\frac{2d+1}{2}  }{(\hat{H}')^d} .\label{centralx}
\end{equation}
The factor of $"3"$ in \eqref{centralx} is introduced as a normalisation, to coincide with the standard result of \cite{Brown:1986nw}.
\\
For the case at hand, comparing with the solutions in \eqref{eq:background} and using Poincar\'e coordinates for AdS$_3$, we have
\begin{eqnarray}
& & a(r,\vec{\theta})=\frac{u}{\sqrt{\hat{h}_4 h_8}} r^2,\;\;\; b(r)=\frac{1}{r^4},\;\;\; d=1,\nn\\
& & \det[g_{ij}]= u\sqrt{\frac{\hat{h}_4^{7}}{h_8}} \frac{\sin^2\chi}{\left(4\hat{ h}_4 h_8) + (u')^2\right)^2},\;\;\; \sqrt{e^{-4\Phi} \det[g_{ij} ] a}= \frac{r}{4}\hat{h}_4 h_8\sin\chi,\nn\\
& & \hat{H}= {\cal N}^2 r^2,\;\;\;\; {\cal N}=\pi \text{Vol}(\text{CY}_2) \int_0^{2\pi(P+1)} \hat{h}_4 h_8 d\rho.
\end{eqnarray}
We then obtain,
\begin{equation}
c_{hol}=\frac{3}{2G_N} {\cal N}= \frac{3\pi}{2G_N}\text{Vol}(\text{CY}_2) \int_0^{2\pi(P+1)} \hat{h}_4 h_8 d\rho= \frac{3}{\pi} \int_0^{2\pi(P+1)} h_4 h_8 d\rho, \label{centralhol}
\end{equation}
where in the last equality we have used--see below \eqref{estax},
\begin{equation}
\Upsilon \text{Vol}(\text{CY}_2)= 16 \pi^4,\;\;\;\;\; \hat{h}_4=\Upsilon h_4,\;\;\; G_N=8\pi^6.\nn
\end{equation}
  
 It is useful to express the holographic central charge in terms of the constants $\alpha_k,\beta_k,\mu_k,\nu_k$ defining the solution,
\begin{equation}
c_{hol}= \sum_{j=0}^P \Bigl(6 \alpha_j \mu_j +3 \alpha_j \nu_j +3\beta_j \mu_j +2\beta_j \nu_j\Bigr).\label{centralholfinal}
\end{equation}
We shall come  back to these expressions in section \ref{ejempl} when we discuss the matching between the holographic quantities studied in this section and the field theory observables discussed below.

\section{The ${\cal N}=(0,4)$ SCFTs }\label{CFTsect}
As we advanced in the Introduction, 
the idea of this work is to propose a duality between the new background solutions in massive IIA found in \cite{Lozano:2019emq} (summarised in  section \ref{geometria})  and 
a set of CFTs. These CFTs are thought to be arising as  low energy fixed points in the RG flows of well defined ${\cal N}=(0,4)$ two dimensional quantum field theories. 
\\
In this 
section we discuss the  weakly coupled UV description of such quantum field theories. \subsection{The UV description}\label{sectionuv}
Let us start with a brief discussion of the fields involved in the weakly coupled description.
It is usual to describe  ${\cal N}=(0,4)$ SUSY in terms of ${\cal N}=(0,2)$
superfields. In this paper we will not use the detailed structure of each $(0,2)$ multiplet. We shall content ourselves with listing the degrees of freedom together with the
R-charges for the fermions involved. As we explain below, these are the details we need to discuss cancellation of gauge anomalies, the R-charge anomaly and the central charge of the IR CFT.
\\
The superfields of ${\cal N}=(0,2)$ two-dimensional SUSY are well described in various references. We found particularly clear and enlightening the papers
\cite{Witten:1993yc}-\cite{Franco:2015tna}.
They contain some of the results we summarise in this section.

As we advanced, instead of going into the details of the $(0,2)$ supermultiplets we describe the degrees of freedom involved in each of them:
\begin{itemize}
\item{\underline{Vector multiplet, $U$}: It contains a gauge field $A_\mu$ and one left moving fermion $\lambda_-$.}
\item{\underline{Chiral multiplet, $\Phi$ }: It consists of a complex scalar $\varphi$ and a right moving fermion $\psi_+$.  By the context, we hope the reader will be able to distinguish between the chiral multiplet and the dilaton in massive IIA, that we denote with the same character $\Phi$.}
\item{\underline{Fermi multiplet, $\Theta$}: This is a constrained superfield for which only a left handed fermion $\psi_-$ propagates. The constraint defining the Fermi superfield generates interactions between the Fermi and the chiral multiplets. The field strength multiplet is an example of a Fermi multiplet. It being constrained agrees with the fact that in two dimensions, a gauge field has no propagating degrees of freedom.}
\end{itemize}
We are interested in theories for which the amount of SUSY is ${\cal N}=(0,4)$. In this case the quantum field theories are formulated in terms
of combinations of $(0,2)$ superfields. For $(0,4)$ SUSY we have:
\begin{itemize}
\item{\underline{$(0,4)$ vector multiplet}: It is expressed as a combination of a $(0,2)$ vector multiplet and a $(0,2)$ Fermi multiplet. There are
two left handed fermions $\lambda_-^{a}$ with $a=1,2$ and a gauge field $A_\mu$. }
\item{\underline{$(0,4)$ hypermultiplet}: Defined as the combination of two chiral multiplets. The degrees of freedom are two complex scalars and two right handed fermions $\psi_+^{a}$.}
\item{\underline{$(0,4)$ twisted hypermultiplet}: Also written as a superposition of two chiral multiplets. The degrees of freedom are two right handed fermions $\tilde{\psi}_+^{a}$ and two complex scalars. The difference with the (non-twisted) hypermultiplet discussed above is in the R-charge assignment. This is reflected in the interactions with other multiplets.}
\item{\underline{$(0,4)$ Fermi multiplet}: It is the superposition of two $(0,2)$ Fermi multiplets. As such, it contains two left handed fermionic degrees of freedom, $\psi_-^{a}$.}
\item{\underline{$(0,2)$ Fermi multiplet}: As explained in \cite{Witten:1994tz}, it is compatible with $(0,4)$ SUSY to have the single left handed fermion of the $ (0,2)$ Fermi multiplet. }

\end{itemize}
The couplings between these multiplets and the constraints on some of them determine the interactions. These can be derived from a superpotential.
See \cite{Witten:1993yc}- \cite{Tong:2014yna} for the details.
\\
In a similar vein one can write the ${\cal N}=(4,4)$ SUSY field content in terms of ${\cal N}=(0,4)$ fields. Notice that in both $(0,4)$ hypers, we have right handed fermions and in the $(0,4)$ vector multiplet left handed ones . In fact, a $(4,4)$ vector multiplet contains a $(0,4)$ vector multiplet and a $(0,4)$ twisted-hypermultiplet (this is: a vector, a Fermi and two chirals of (0,2) SUSY). A ${\cal N}=(4,4)$ hypermultiplet contains a $(0,4)$ hypermultiplet and a $(0,4) $ Fermi multiplet, hence containing two Fermi and two chiral multiplets of $(0,2)$ SUSY.

The R-symmetry of ${\cal N}=(0,4)$ field theories is SU(2)$_L\times$SU(2)$_R$. We single out a U(1$)_R$ inside SU(2)$_R$ and quote the U(1)$_R$ charge of each fermion in the above multiplets. This will be used below to calculate the anomaly of the global R-symmetry. See equation (3.13) in the paper \cite{Putrov:2015jpa} for the same charge assignment.\\
 For the $(0,4)$ vector multiplet we have that the left handed fermion inside the vector has $R[\lambda_-^v]=0$ while the left handed fermion inside the Fermi multiplet has $R[\lambda_-^f]=1$. Similarly, for the $(0,4)$ twisted hypermultiplet we have that for both right handed fermions $R[\tilde{\psi}_+^a]=0$. For both right handed fermions inside the $(0,4)$  hypermultiplet we have $R[{\psi}_+^{a}]=-1$. Finally, the fermion inside the $(0,2)$ Fermi multiplet (allowed in theories with $(0,4)$ SUSY) is such that $R[\lambda_-^f]=0$.
\\
Now, we explore the condition for cancellation of gauge anomalies.
\subsection{Anomaly cancellation}
We are dealing  with chiral theories. Their consistency requires one to be careful with the field content, so that gauge anomalies are vanishing.
In this work we only need to use that the anomaly of a (gauged or global) non-Abelian symmetry is given by the correlator of the symmetry currents, $<J_\mu^A(x) J_\nu^B(x)>\sim k \delta^{A,B}  \delta_{\mu\nu}$. Notice that there is no mixing between non-Abelian currents. On the other hand, Abelian currents can mix. The coefficient $k$ is calculated by computing  $\Tr [\gamma_3 J_{\text{SU}(N)} J_{\text{SU}(N)}  ]$. This should be read as the difference between the right handed fermions times their charge squared and the left handed fermions times their charge squared. Let us study in detail the contribution  to the SU($N$) anomaly coming from the various ${\cal N}=(0,2)$ multiplets mentioned above:
\begin{itemize}
\item{Chiral multiplets: If they are in the adjoint representation of the symmetry group SU($N$), they contribute with a factor $N$. If they transform in the (anti) fundamental, they contribute with a factor $\frac{1}{2}$.}
\item{Fermi multiplets: If they are in the adjoint representation of the symmetry group SU($N$), they contribute with a factor $-N$. If they transform in the (anti) fundamental, they contribute with a factor $-\frac{1}{2}$.}
\item{Vector multiplets: They are in the adjoint representation of the symmetry group SU($N$). They contribute with a factor $-N$. }
\end{itemize}
\subsection{Building block of our theories}
\begin{figure}[h!]
    \centering
    {{\includegraphics[width=5cm]{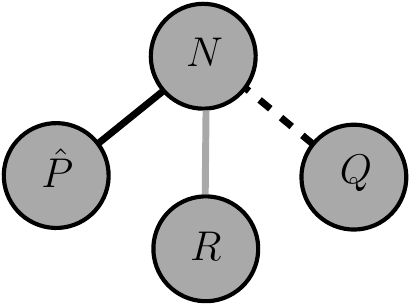} }}%

\caption{The building block of our theories. The solid black line represents a $(4,4)$ hypermultiplet. The grey line represents a $(0,4)$ hypermultiplet. The dashed line represents a $(0,2)$ Fermi multiplet. Inside the gauge group SU($N$) run $(4,4)$ SUSY vector multiplets. The groups SU($\hat{P}$), SU($Q$) and SU($R$) can be gauge or global.}
\label{explanation}

\end{figure}

Let us discuss now what will be the 'building block' of our quantum field theories. See figure \ref{explanation}.
We have an SU($N$) gauge group. In the gauge group the matter content is that of a $(4,4)$ vector multiplet, namely---in $(0,2)$ notation,  a vector, two twisted chirals and a Fermi multiplet in the adjoint representation of SU($N$). This gauge group is joined with other (gauged of global) symmetry groups SU($\hat{P}$), SU($R$) and SU($Q$). The connection with the SU($\hat{P}$) symmetry group is mediated by  $(4,4)$ hypers. In $(0,2)$ notation, $2\times N\times \hat{P}$ Fermi multiplets and $2\times N\times \hat{P}$ chiral multiplets run over the black solid line. The connection with the SU($R$) symmetry group is via   $(0,4)$ hypermultiplets. In $(0,2)$ notation
$2\times N\times R$ chiral multiplets propagate over the grey lines. Finally, over the dashed line run $N \times Q$ Fermi multiplets in $(0,2)$ notation. Notice that  a similar (but not the same!) field content to this was proposed in \cite{Gadde:2015tra},  in the study  of the field theories associated with tensionless strings in ${\cal N}=(0,1)$ six-dimensional SCFTs.
\\
Let us now calculate the anomaly of the gauged SU($N$) symmetry group and impose that it vanish. We focus only on the gauged SU($N$) group, but a similar job should be done for all other gauged symmetry groups. Let us spell the various contributions:
\begin{itemize}
\item{The contribution of the adjoint fields is $2N-N-N=0$. This is expected, as the field content is that of a $(4,4)$ vector multiplet.}
\item{The contribution of the bifundamentals connecting with $\text{SU}(\hat{P})$ is $(\frac{1}{2}-\frac{1}{2})2 \hat{P} N=0$. Again, this vanishing contribution is expected as we are dealing with  $(4,4)$ hypers.}
\item{The link with the symmetry $\text{SU}(R)$ contributes a factor $2\times N\times R\times \frac{1}{2}= N R$.}
\item{Finally the bifundamentals running on the link with the $\text{SU}(Q)$ symmetry group contribute $-\frac{1}{2} N Q$.}
\end{itemize}
Thus, in order to have a non anomalous gauged symmetry we need to impose that the four contributions above add to zero, that is
\begin{equation}
2R=Q \, .\label{noanomaly}
\end{equation}
This mechanism should apply to all other gauged symmetry groups. When we construct our gauge theories, they will be represented by quivers obtained by 'assembling' the building blocks of figure \ref{explanation}.
\subsection{U(1) R-symmetry anomaly}
It is instructive to compute the R-symmetry anomaly for our 'building block'. Once again, we focus the attention on the $\text{SU}(N)$ gauge group. We use the values for the U(1)$_R$ charges quoted near the end of section \ref{sectionuv}. We find that the U(1)$_R$ anomaly, following from $\Tr[\gamma_3 Q_i^2]$ is given by the sum of various contributions. In detail, we have,
\begin{itemize}
\item{For the fields in the adjoint of  the $\text{SU}(N)$ gauge group, the only contribution is from the fermions inside the Fermi multiplet (all the other fermions have zero $U(1)_R$ charge). The contribution of these particular left handed fermions is $- (N^2-1)$. This coincides with (minus) the number of $(0,4)$ vector multiples in $\text{SU}(N)$. }
\item{The contribution from the bifundamentals joining $\text{SU}(N)$ with $\text{SU}(\hat{P})$ is $N\times \hat{P}$. This is the number of $(0,4)$ hypermultiplets in that link.}
\item{The contribution coming from the fields running over the grey line, joining $\text{SU}(N)$ with $\text{SU}(R)$, is $N\times R$, once again, counting the number of $(0,4)$ hypers running on  the link.}
\item{Finally, the fields running over the dashed line do not contribute as the R-charge of the left handed fermion is zero, as we discussed above.}
\end{itemize}
In summary, we find that
\begin{equation}
\Tr[\gamma_3 Q_i^2]\sim (n_{hyp}- n_{vec}).\label{r-anomaly}
\end{equation}
Thus, the R-symmetry anomaly is proportional to the number of $(0,4)$ hypers minus the number of $(0,4)$ vectors.

\subsection{Central charge, R-anomaly and the superconformal algebra}
Up to this point, we have found the condition for our building block to be non-anomalous, see \eqref{noanomaly}, and the contribution of the matter charged under $\text{SU}(N) $ to the U(1)$_R$ anomaly, see \eqref{r-anomaly}. If the theory becomes conformal and strongly coupled -- as we shall propose our quivers do when flowing to low energies -- the coefficients for the anomalies cannot be computed by summing over fermions at the conformal point (as we do not have a particle-like description of the CFT). But since these  coefficients are 't Hooft anomalies, they are invariants under RG-flow.  Hence UV-QFT calculations are good for the same IR-CFT quantity (we are assuming that the proposed R-symmetry does not mix in the IR with other Abelian symmetries). We  propose that our quivers become conformal in the IR and then the central charge of the quiver and the R-symmetry anomaly get related by the superconformal algebra.

In our case the relevant superconformal algebra is the small ${\cal N}=(0,4)$  algebra. This consists of eight operators: the energy momentum tensor $T(z)$, four fermionic superpartners $G^a(z)$ and three Kac-Moody currents $J^{i} (z)$. The dimensions of these operators are $(2, \frac{3}{2},1)$ respectively. The modes of these operators satisfy an algebra that can be derived from the OPE's of the small ${\cal N}=(0,4)$ algebra. In particular among the various relations we have,
\begin{eqnarray}
& & T(z)T(0)\sim \frac{c}{z^4}+ 2\frac{T(0)}{z^2} +\frac{\partial T}{z}+\text{regular},\;\;\; J^i(z)J^l(0)\sim \frac{k^{il}}{z^2}+\text{regular}.\nn
\end{eqnarray}
A relation between $c$ and $k^{il}=k \delta^{il}$ appears by virtue of the algebra of (anti) commutators. The relation is that $c= 6\times k$. In other words, for our building block
\begin{equation}
c= 6 (n_{hyp}- n_{vec}).\label{important}
\end{equation}
This relation---also derived in \cite{Putrov:2015jpa}, is of importance to us. Let us briefly discuss it, as well as the proposed duality and its implications.
\subsection{The proposed duality}
\begin{figure}[h!]
    \centering
    {{\includegraphics[width=10cm]{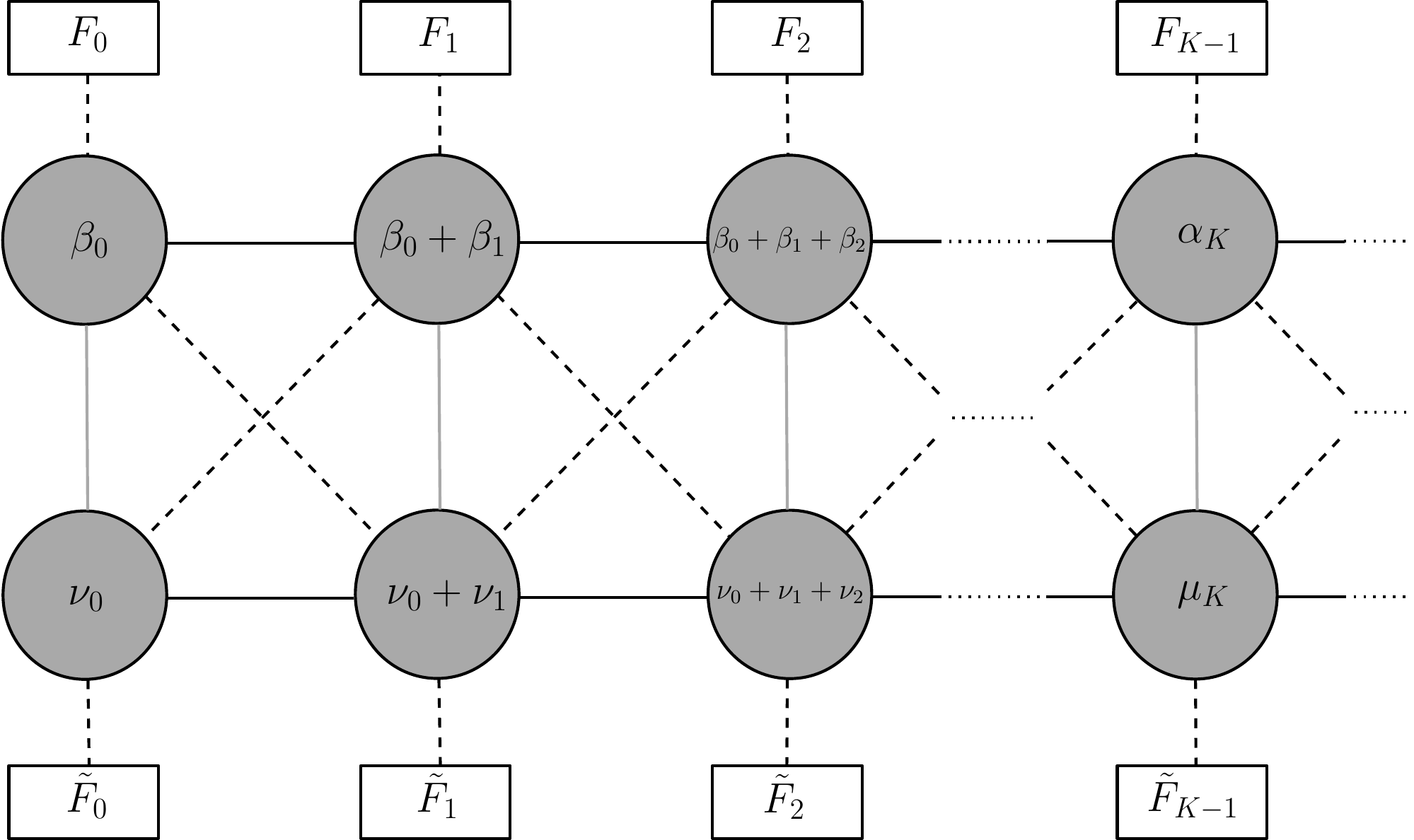} }}%

\caption{ A generic quiver field theory whose IR is dual to the holographic background defined by the functions in \eqref{profileh4sp}-\eqref{profileh8sp}. As before, the solid black line represents a $(4,4)$ hypermultiplet. The grey line represents a $(0,4)$ hypermultiplet and  the dashed line represents a $(0,2)$ Fermi multiplet. ${\cal N}=(4,4)$ vector multiplets are the degrees of freedom  in each gauged node.}
\label{figurageneral}

\end{figure}

\begin{figure}[h!]
    \centering
    {{\includegraphics[width=11cm]{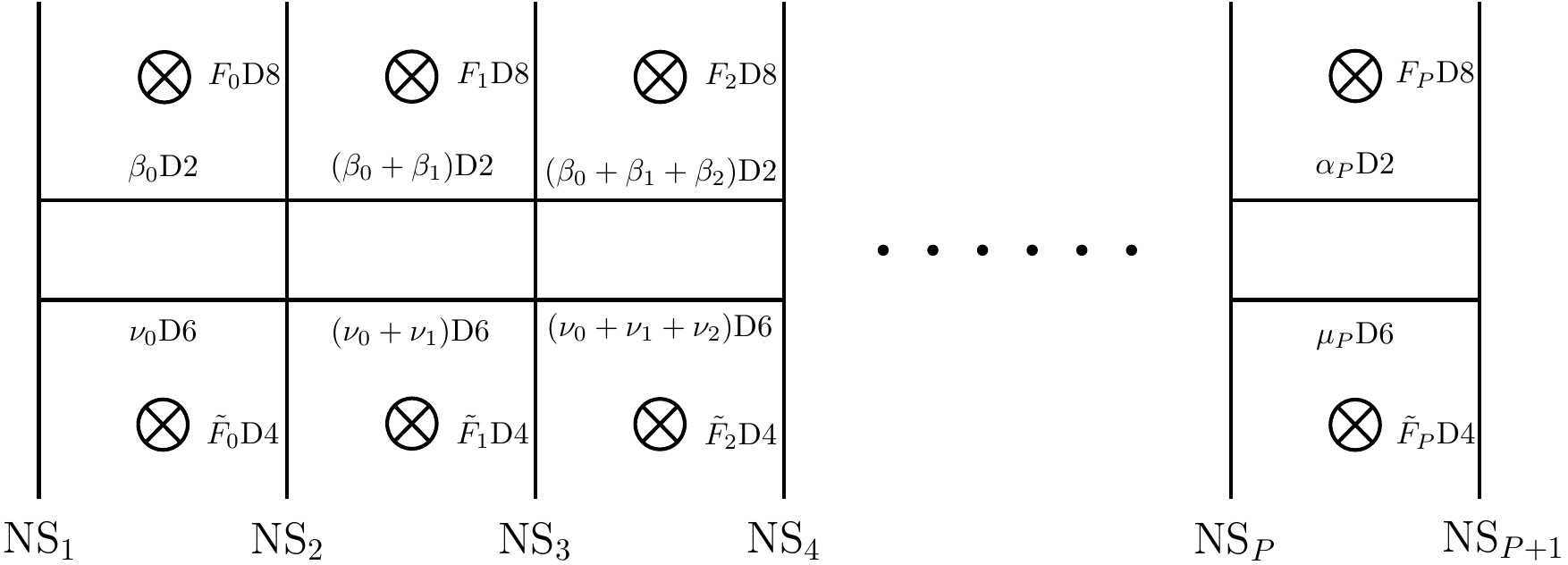} }}%

\caption{Hanany-Witten set-up associated with our generic quiver in figure \ref{figurageneral}. The vertical lines denote NS five branes, horizontal lines D2 and D6 colour branes. The crosses, D4 and D8 flavour branes. }
\label{vvvbb}

\end{figure}

In what follows we shall define ${\cal N}=(0,4)$ SUSY quiver field theories. These quivers will consist of colour and flavour groups joined by hypermultiplets or ${\cal N}=(0,2)$ Fermi multiplets as indicated in our building block. We must be careful to have all anomalies of gauged groups vanishing. We will also calculate the R-symmetry anomaly and the 'central charge' via the relation in \eqref{important}\footnote{Strictly speaking, we should not call this quantity central charge as (in the UV) we are not at a fixed point of the RG flow. The relation in \eqref{important} is only valid at the fixed point. }. The calculation will be performed in the weakly coupled description of the field theory, in the UV before the conformal point is attained. But as we mentioned, these are 't Hooft coefficients, hence invariants of the RG flow. Importantly, we assume that there is no mixing between the R-symmetry and other global symmetries. If such mixing were to exist, an extremisation procedure like the one devised in \cite{Benini:2012cz},\cite{Benini:2013cda} would be needed. It would be nice to prove that for our quivers there is no mixing between the R-symmetry and other global symmetries. As a plausible  argument for the non-mixing, notice that the non-Abelian R-symmetry $SU(2)$ cannot mix with $U(1)$ global symmetries in two dimensions. There is no other non-Abelian global R-symmetry to mix with. Let us then focus on the end of the RG flow to low energies.
\\
As advanced, we propose that our quivers flow to a strongly coupled CFT with ${\cal N}=(0,4)$ SUSY and central charge given  by \eqref{important}, as enforced by the superconformal algebra.  The second part of our proposal is that the holographic backgrounds are dual to these CFTs. The holographic central charge calculated in \eqref{centralhol} should coincide with the result of \eqref{important}, in the case of long quivers with large ranks (as this is the regime in which we can trust the supergravity solutions).
\\
Another check of our proposal will be the matching of global symmetries on both sides of the duality. In fact the SCFTs have $SO(2,2)$ space-time  and SU(2) R-symmetries. The backgrounds in \eqref{eq:background} match these with the isometries of AdS$_3$ and S$^2$ respectively.  Also eight  supercharges are preserved both by the CFT and the background. Indeed, there are four space-time (Q's) and four conformal (S's) supercharges. More interestingly, the flavour symmetries of the SCFT are matched by the presence of 'flavour branes'  in the background (giving place to Bianchi identities modified by the presence of sources).  The counting of Page charges also coincides with the ranks of the colour and flavour groups, or, analogously, with the numbers of (D2,D6) colour branes and (D4,D8) flavour branes in the associated Hanany-Witten brane set-ups.
\\
Let us be more concrete. A generic background of the form in \eqref{eq:background} is defined by the functions $\hat{h}_4,h_8, u$. In the type of solutions we consider in this paper (those where the space ends at $\rho_*=2\pi(P+1)$, where we have $\hat{h}_4(\rho_*)=h_8(\rho_*)=0$),  we generically have---see \eqref{profileh4final}-\eqref{profileh8final} and \eqref{bati},
 \begin{equation} \label{profileh4sp}
\hat{h}_4(\rho)\!=\!\Upsilon\! h_4(\rho)\!=\!\!
                    \Upsilon\!\!\left\{ \begin{array}{ccrcl}
                       \frac{\beta_0 }{2\pi}
                       \rho & 0\leq \rho\leq 2\pi \\
                 \beta_0+\frac{\beta_1}{2\pi}(\rho-2\pi)      &2\pi \leq \rho \leq 4\pi \\
(\beta_0+\beta_1 )+\frac{\beta_2}{2\pi} (\rho-4\pi) & 4\pi\leq\rho\leq 6\pi\\
                                     (\beta_0+\!\beta_1\!+\!....+\!\beta_{k-1})\! +\! \frac{\beta_k}{2\pi}(\rho-2\pi k) &~~ 2\pi k\leq \rho \leq 2\pi(k+1),\;\;\;\; k:=3,....,P-1\\
                      \alpha_P-  \frac{\alpha_P}{2\pi}(\rho-2\pi P) & 2\pi P\leq \rho \leq 2\pi(P+1).
                                             \end{array}
\right.
\end{equation}
 \begin{equation} \label{profileh8sp}
h_8(\rho)
                    =\left\{ \begin{array}{ccrcl}
                       \frac{\nu_0 }{2\pi}
                       \rho & 0\leq \rho\leq 2\pi \\
                 \nu_0+\frac{\nu_1}{2\pi}(\rho-2\pi)      &2\pi \leq \rho \leq 4\pi \\
(\nu_0+\nu_1 )+\frac{\nu_2}{2\pi} (\rho-4\pi) & 4\pi\leq\rho\leq 6\pi\\
                                     (\nu_0+\nu_1+....+\nu_{k-1}) + \frac{\nu_k}{2\pi}(\rho-2\pi k) &~~ 2\pi k\leq \rho \leq 2\pi(k+1),\;\;\;\; k:=3,....,P-1\\
                      \mu_P-  \frac{\mu_P}{2\pi}(\rho-2\pi P) & 2\pi P\leq \rho \leq 2\pi(P+1).
                                             \end{array}
\right.
\end{equation}
and \begin{equation}
u=\frac{b_0}{2\pi}\rho.\nn
\end{equation}
The background in \eqref{eq:background} for the functions  $\hat{h}_4,h_8, u$ above is dual to the CFT describing the low energy dynamics of a two dimensional quantum field theory encoded by the  quiver in figure \ref{figurageneral} and the Hanany-Witten set-up of figure \ref{vvvbb}.

Let us see how the correspondence works. For the first two gauge groups $\text{SU}(\nu_0)$ and $\text{SU}(\beta_0)$, the cancellation of gauge anomalies in \eqref{noanomaly} implies that,
\begin{equation}
F_0+\nu_0+\nu_1=2\nu_0\to F_0=\nu_0-\nu_1,\;\;\; \tilde{F}_0+\beta_0+\beta_1=2\beta_0\to \tilde{F}_0=\beta_0-\beta_1.
\end{equation}
This is precisely the number of flavour D8 and D4 branes predicted by the Bianchi identities in the interval $[0,2\pi]$---see
\eqref{numberofbranes} for $k=1$. Similarly, the ranks of the first two gauge groups, namely $\beta_0$ and $\nu_0$, are precisely the numbers of D2 and D6 colour branes predicted by eq.\eqref{numberofcolour} in the first interval (for $k=1$).
\\
This works similarly for all other entries in the quiver. For example, for the $\text{SU}(\alpha_k)$ colour group, we obtain that in the interval $[2\pi(k-1), 2\pi k]$ of the associated Hanany-Witten set up in figure \ref{vvvbb}, there are $\alpha_k$ D2 branes, with 
\begin{equation}
\alpha_k= \sum_{j=0}^{k-1} \beta_j.\nn
\end{equation}
The cancellation of gauge anomalies for the $\text{SU}(\alpha_k)$ gauge group imposes that,
\begin{equation}
F_{k-1}+\mu_{k+1}+\mu_{k-1}=2\mu_k\to  F_{k-1}=\nu_{k-1}-\nu_k,
\end{equation}
which, according to \eqref{numberofbranes},  is precisely the number of flavour D8 branes in the $[2\pi(k-1), 2\pi k]$ interval of the brane set-up. Things work analogously if we replace D2 for D6  (or $ \alpha_k\to\mu_k$) and D8 for D4 ($\nu_k\to \beta_k$) and deal with the lower-row gauge group $\text{SU}(\mu_k)$.
\\
We can calculate the field theory central charge by counting the number of $(0,4)$ hypermultiplets,  the number of $(0,4)$ vector multiplets and using \eqref{important}. We find,
\begin{eqnarray}
& & n_{vec}= \sum_{j=1}^P \Bigl(\alpha_j^2+\mu_j^2-2\Bigr),\;\;\;\; n_{hyp}= \sum_{j=1}^P \alpha_j\mu_j +\sum_{j=1}^{P-1} \Bigl(\alpha_j \alpha_{j+1}+\mu_j\mu_{j+1}\Bigr),\nn\\
& & c= 6\times \left(   \sum_{j=1}^P \Bigl(\alpha_j\mu_j -\alpha_j^2-\mu_j^2+2\Bigr) +\sum_{j=1}^{P-1} \Bigl(\alpha_j \alpha_{j+1}+\mu_j\mu_{j+1}\Bigr)  \right).\label{batigol}
\end{eqnarray}
When the number of nodes is large $P>>1$, and the ranks of each gauge group $\alpha_i,\mu_i$ are large numbers, the supergravity backgrounds are trustable and the holographic central charge calculated according to \eqref{centralhol} should coincide at leading order in these large parameters  with \eqref{batigol}.
\\
For pedagogical purposes, in the next section we present some explicit examples (in increasing  level of complexity) of quiver-supergravity dual pairs. We shall check the cancellation of gauge anomalies and the leading order matching of \eqref{centralhol} and \eqref{batigol}.

\section{Various checks of our proposed duality}\label{ejempl}
In this section we discuss various examples of 
dual holographic  pairs. We check anomaly cancellation and the leading order matching of the CFT and holographic central charges. We start from
the simplest possible example of a quiver field theory flowing to a superconformal ${\cal N}=(0,4)$ SCFT that admits a viable supergravity dual, and move on to examples of increasing complexity. These will provide stringent checks of our proposal\footnote{In the examples that follow we write the function $h_4(\rho)$. As discussed above, the function that appears in the background is $\hat{h}_4=\Upsilon h_4$. The value $\Upsilon \text{Vol(CY}_2)=16\pi^4$ is used to have well quantised charges in terms of the integer numbers ($\alpha_k,\beta_k,\mu_k,\nu_k$).}.

\subsection{Example I}\label{ejemplos1}
\begin{figure}[h!]
    \centering
    {{\includegraphics[width=11cm]{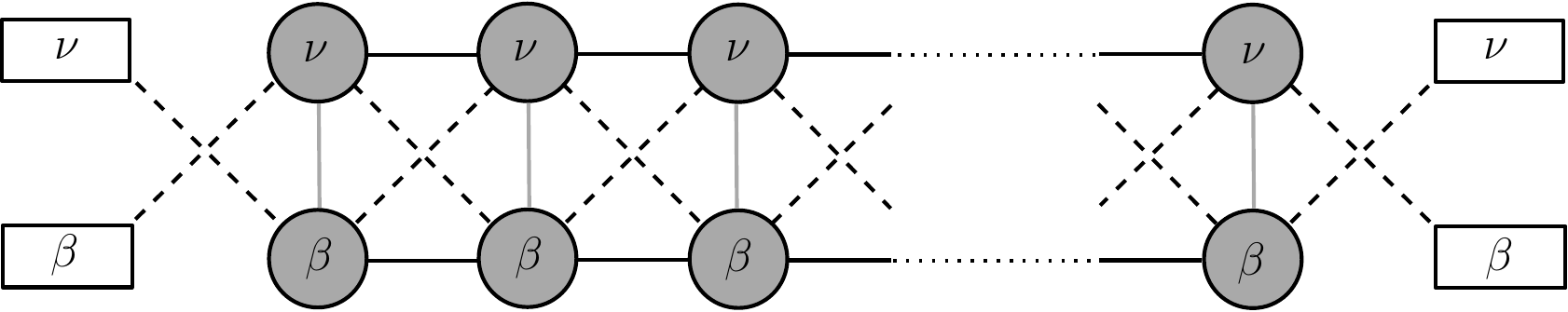} }}%

\caption{ The quiver encoding our first example of quantum field theory. The conventions for the fields running along the different lines are the same as those in section \ref{CFTsect}.}
\label{example1}

\end{figure}

Consider the quiver of figure \ref{example1}, where we depict $P$ gauge groups $\text{SU}(\nu)$ and $P$ gauge groups $\text{SU}(\beta)$. They are joined by bifundamentals, all complemented by flavour groups (rectangular boxes). This quiver encodes the kinematical content of our first field theory. We propose that this QFT flows in the IR to a CFT.
 Let us focus on the first gauge group of the top row, $\text{SU}(\nu)$. We compare with our building block in figure \ref{explanation} to find that,
\begin{eqnarray}
\hat{P}=\nu,\;\;\; Q=2\beta,\;\;\; R=\beta.
\end{eqnarray}
This is precisely what our formula (\ref{noanomaly}) requires  for the cancellation of the $\text{SU}(\nu)$ gauge anomaly.
For the first $\text{SU}(\beta)$ gauge group in the lower row, we have $\hat{P}=\beta, Q=2\nu, R=\nu$ and \eqref{noanomaly} is also satisfied.
\\
 Similarly, one can calculate for the top and bottom  gauge groups at the right end of the figure and check that all of them satisfy \eqref{noanomaly}. Finally, for any intermediate $\text{SU}(\nu)$-node, we have $\hat{P}=\nu$, $Q=2\beta$, $R=\beta$. Analogous statements hold true for the lower row groups. Hence  all of the gauge symmetries are  non-anomalous.
\\
We can now calculate the number of $(0,4)$ hypermultiplets and vector multiplets with a view on computing the central charge of the  IR CFT. We find,
\begin{eqnarray}
& & n_{vec}=P(\nu^2+\beta^2-2),\;\;\;\; n_{hyp}= (P-1)(\nu^2+\beta^2) + P\nu\beta.\nonumber\\
& & c=6(n_{hyp}-n_{vec})= 6\nu\beta P(1 +\frac{2}{\beta\nu} -\frac{\beta}{\nu P}-\frac{\nu}{\beta P})\sim 6\nu\beta P.\label{centralQFTI}
\end{eqnarray}
In the last approximation we used that the ranks are large numbers $(\nu,\beta)\to \infty$ and that the quiver is long enough, hence $P>>1$, to meaningfully compare with the dual massive IIA solution.
\\
The holographic background dual to this CFT is given in terms of the functions $u=\frac{b_0}{2\pi}\rho$ and
\begin{equation} \label{profileh8exampleII}
h_8(\rho)
                    =\left\{ \begin{array}{ccrcl}
                       \frac{\nu }{2\pi}
                       \rho & 0\leq \rho\leq 2 \pi \\
                 \nu      &2\pi \leq \rho \leq 2\pi P \\
                      \frac{\nu }{2\pi}(2\pi( P+1) -\rho) & 2\pi P\leq \rho \leq 2\pi(P+1).
                                             \end{array}
\right.
\end{equation}
  \begin{equation} \label{profileh4exampleII}
h_4(\rho)
                    =\left\{ \begin{array}{ccrcl}
                       \frac{\beta }{2\pi}
                       \rho & 0\leq \rho\leq 2 \pi \\
                 \beta     &2\pi \leq \rho \leq 2\pi P\\
                      \frac{\beta }{2\pi}(2\pi( P+1) -\rho) & 2\pi P\leq \rho \leq 2\pi(P+1).
                                             \end{array}
\right.
\end{equation}
The holographic central charge is found by the simple calculation in \eqref{centralhol},
\begin{eqnarray}
& & c_{hol}= \frac{3}{\pi} \left(\int_0^{2\pi} \frac{\beta\nu}{4\pi^2}\rho^2 d\rho +\int_{2\pi}^{2\pi P}\beta\nu d\rho +\int_{2\pi P}^{2\pi (P+1)} \frac{\beta\nu}{4\pi^2} (2\pi( P+1) -\rho)^2d\rho   \right)\nn\\
& & c_{hol}=6\beta \nu P (1-\frac{1}{3P})\sim 6 P \beta\nu.
\end{eqnarray}
This coincides with the field theoretical result in \eqref{centralQFTI}.
Finally, notice that the number of D4 and D8 flavour branes, dictated by \eqref{numberofbranes}, precisely provide the flavour symmetries at the beginning and end of the quiver. One finds the same by inspecting \eqref{numberofcolour} for the number of colour branes, coinciding with the ranks of the gauge groups of our quiver.

\subsection{Example II}\label{ejemplos2}
\begin{figure}[h!]
    \centering
    {{\includegraphics[width=10cm]{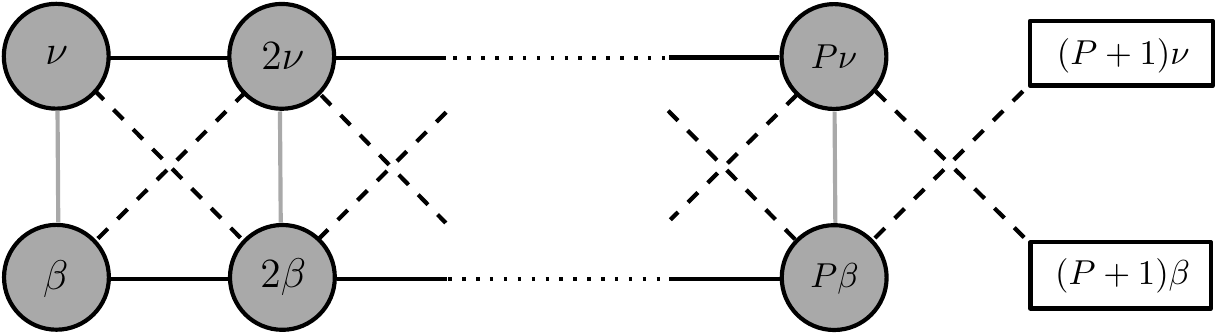} }}%

\caption{ The quiver encoding our second example. There are $P$ gauged nodes with increasing rank in each row. The conventions for the fields running along the different lines are the same as those in section \ref{CFTsect}.}
\label{segundo}

\end{figure}

Let us slightly complicate our previous example. We consider now a quiver with two rows of linearly increasing colour groups. These two rows are finished after $P$ nodes by the addition of a flavour group for each row. See figure \ref{segundo}.
This type of quivers can be used as a completion of the background obtained via the application of non-Abelian T-duality on AdS$_3\times $S$^3\times$CY$_2$, inspired by the treatments in \cite{Lozano:2016kum}-\cite{Lozano:2018pcp}.  See \cite{LMNR3} for a careful discussion of this.
The anomalies of each of the gauge groups can be easily seen to vanish. In fact, for any of the intermediate gauge nodes, say SU($k \nu$) and referring to our building block in figure \ref{explanation}, we have $ Q= 2 k\beta , R=k\beta$. This implies that \eqref{noanomaly} is satisfied and a generic intermediate gauge group is not anomalous. If we refer to the last gauge group in the upper-row  SU($P\nu$) we have that $Q=(P+1)\beta+(P-1)\beta=2P\beta$ and $R=P\beta$. As a consequence  \eqref{noanomaly} is satisfied and the gauged group SU($P\nu$)  is not anomalous. The same occurs for the lower-row gauge groups.
\\
We can easily count the number of $(0,4)$ hypers and the number of $(0,4)$ vector multiplets,
\begin{eqnarray}
 n_{vec}=\sum_{j=1}^P \Bigl(j^2(\nu^2+\beta^2)-2\Bigr),\;\;\;
 n_{hyp}=\sum_{j=1}^{P-1} j(j+1)(\nu^2+\beta^2)+ \sum_{j=1}^{P} j^2\nu\beta.
\end{eqnarray}
The central charge of the IR CFT is,
\begin{eqnarray}
c&=&6(n_{hyp}-n_{vec})\nonumber\\
&=&6\nu\beta (\frac{P^3}{3} +\frac{P^2}{2} +\frac{P}{6}) - 3(\nu^2+\beta^2)(P^2+P)+12 P\sim 2\nu\beta P^3.\label{centralchargeexampleII}
\end{eqnarray}
The holographic description of this system is in terms of the functions,
\begin{equation} \label{profileh8exampleII}
h_8(\rho)
                    =\left\{ \begin{array}{ccrcl}
                       \frac{\nu }{2\pi}
                       \rho & 0\leq \rho\leq 2\pi P \\
                      \frac{\nu P}{2\pi}(2\pi( P+1) -\rho) & 2\pi P\leq \rho \leq 2\pi(P+1).
                                             \end{array}
\right.
\end{equation}
  \begin{equation} \label{profileh4exampleII}
h_4(\rho)
                    =\left\{ \begin{array}{ccrcl}
                       \frac{\beta }{2\pi}
                       \rho & 0\leq \rho\leq 2\pi P\\
                      \frac{\beta P}{2\pi}(2\pi (P+1) -\rho) & 2\pi P\leq \rho \leq 2\pi(P+1).
                                             \end{array}
\right.
\end{equation}
Using \eqref{centralhol}, we calculate the holographic central charge,
\begin{equation}
c_{hol}=\frac{3}{\pi} \left(\frac{\beta\nu}{4\pi^2}\right) \left(\int_0^{2\pi P}\rho^2 d\rho+ \int_{2\pi P}^{2\pi(P+1)}P^2  (2\pi (P+1) -\rho)^2 d\rho    \right)=2\nu\beta P^3(1+\frac{1}{P})\sim 2 \nu\beta P^3.\label{ccholii}
\end{equation}
Again, we observe that in the limit of a long quiver, there is matching for the central charge in the CFT --see \eqref{centralchargeexampleII},  with that of the dual description--see \eqref{ccholii}.
\\
Let us now discuss a  more involved  example, providing us with a much  stringent check of our proposed duality. 
\subsection{Example III}
\begin{figure}[h!]
    \centering
    {{\includegraphics[width=14cm]{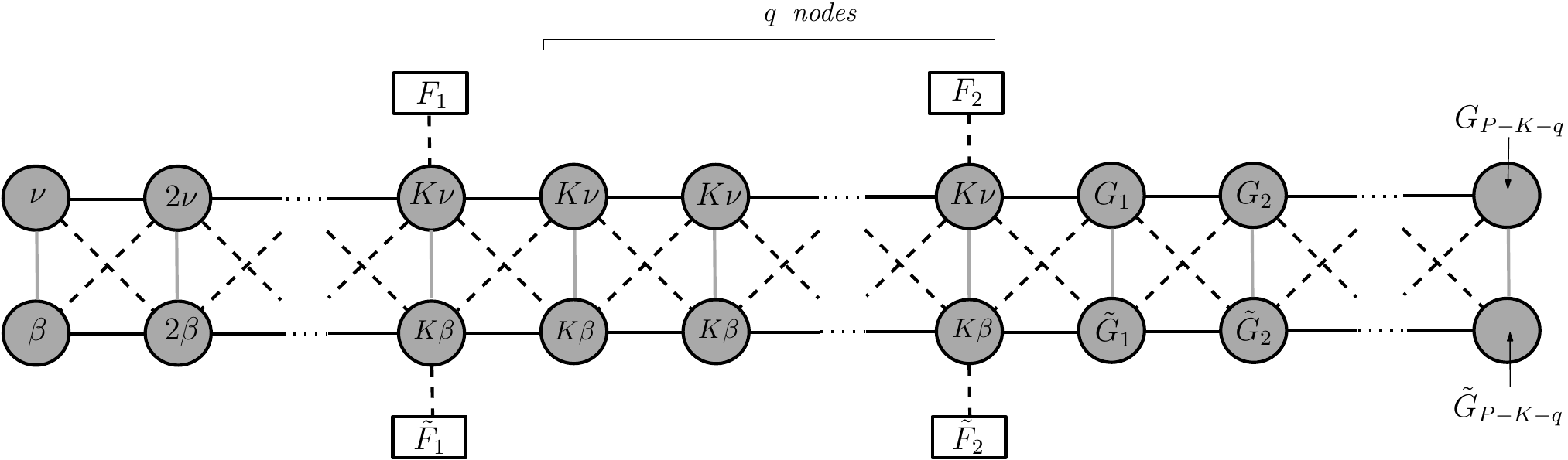} }}%

\caption{ The quiver encoding our third example. There are $K$ gauged nodes with linearly increasing ranks in each row.  These are followed by $q- \text{SU}(K\nu)$ (top row) and $q-\text{SU}(K\beta)$ nodes (lower row). The ranks of the next $\text{SU}(G_i)$ and $\text{SU}(\tilde{G}_i$) nodes is given in the text. The conventions for the fields running along the different lines are the same as those in section \ref{CFTsect}.}
\label{tercera}

\end{figure}

In this case we consider the more involved field theory encoded by the quiver in figure \ref{tercera}.

In this quiver we have a line of linearly increasing nodes $\text{SU}(\nu)\times \text{SU}(2\nu)\times ....\times \text{SU}(K\nu)$ followed by $q\times \text{SU}(K\nu)$ nodes. The gauge groups $\text{SU}(G_l)$ have ranks 
\begin{equation}
G_l=K\nu(1-\frac{l}{P+1-K-q}),\;\;\;\;\; l=1,...., P-K-q. \label{gls}
\end{equation}
 For the lower row we have analogous kinematics: Linearly increasing ranks $\text{SU}(\beta)\times \text{SU}(2\beta)\times ....\times \text{SU}(K\beta)$, followed by $q\times \text{SU}(K\beta)$ nodes. The gauge groups $\text{SU}(\tilde{G}_l)$ have ranks, 
\begin{equation}
\tilde{G}_l=K\beta(1-\frac{l}{P+1-K-q}),\;\;\;\;\; l=1,...., P-K-q.\label{tgls}
\end{equation}

Let us analyse anomalies for the upper row groups (the lower row ones work analogously). The linearly increasing chain is non-anomalous like our previous example in section \ref{ejemplos2} was. Namely, for a generic $\text{SU}(j \nu)$ node, we have $Q=2 j \beta$ and $R=j\beta$. \\
The chain of $q$ $\text{SU}(K\nu)$ groups works exactly as any intermediate group in section \ref{ejemplos1}, namely for any generic (intermediate) node we have $Q=2 K\beta$ and $R=K\beta$, satisfying \eqref{noanomaly}. \\
More interesting are the first and last of these $q$-nodes. For the first node we have $Q= F_1+(K-1)\beta +K \beta$ and $R= K\beta$. Observe that \eqref{noanomaly} forces 
\begin{equation}
F_1= \beta.\nn
\end{equation}
 For the last of these $q$-nodes we have $Q=K\beta + \tilde{G}_1 + F_2$ and $R=K\beta$. Then the vanishing of the gauge anomaly forces 
 \begin{equation}
 F_2= \frac{K\beta}{P+1-K-q}.\nn
 \end{equation}
 For any generic group $\text{SU}(G_i)$ we have $Q= \tilde{G}_{i-1} +\tilde{G}_{i+1}$ and $R=\tilde{G}_i$. Using \eqref{tgls} we find that $Q=2R$ as imposed in \eqref{noanomaly} for the vanishing of the gauge anomalies.
\\
Analogously, for the lower row groups, we find that the vanishing of the gauge anomalies imposes
\begin{equation}
\tilde{F}_1=\nu,\;\;\;\;\; \tilde{F}_2=\frac{K\nu}{P+1-K-q}.\label{baba}
\end{equation}
To calculate the CFT central charge we need to compute the number of $(0,4)$ hypers and vectors. We find
\begin{eqnarray}
n_{vec}&=&\sum_{j=1}^K\left(j^2(\nu^2+\beta^2)-2\right) +  q(K^2 (\nu^2+\beta^2)-2)\nonumber\\
 & &+\sum_{j=1}^{P-K-q} \left(K^2(\nu^2+\beta^2)(1-\frac{j}{P+1-K-q})^2-2 \right)     ,\nonumber\\
 n_{hyp}&=& \sum_{j=1}^{K-1}  j(j+1)(\nu^2+\beta^2) +\sum_{j=1}^{K}j^2 \beta \nu + K^2 q(\nu^2+\beta^2+\beta\nu)\nonumber\\
 & &+\sum_{j=0}^{P-K-q-1} K^2(\beta^2+\nu^2)(1-\frac{j}{P+1-K-q})(1-\frac{j+1}{P+1-K-q})\nonumber\\ 
 & &+\sum_{j=1}^{P-K-q} K^2\beta \nu (1-\frac{j}{P+1-K-q})^2.
 \end{eqnarray}
The field theory central charge is after a lengthy calculation,
\begin{eqnarray} \label{centralchargeexampleIV}
c&=&6(n_{hyp}-n_{vec})\nonumber\\
                    &\sim&\left\{ \begin{array}{ccrcl}
                      2\beta\nu K^2P + 12 P + O(1, 1/P),\;\; &\text{if} ~P>>1,\\
                       4\beta \nu K^2 q + O(1, 1/q),\;\;&\text{if} ~ q>>1,\\
                       2\beta\nu K^2(1+2q+P)+ O(1, 1/K),\;\; &\text{if} ~K>>1.
\end{array}
\right.
\end{eqnarray}
We have expanded the exact result for the three possible ways in which the quiver may be considered to be 'long'. We  also need to take $(\nu,\beta)$ to be large numbers. \\
Now, let us compare with the holographic description. The functions $h_4$ and $h_8$ for this case read,
\begin{equation} \label{profileh8exampleIV}
h_8(\rho)
                    =\left\{ \begin{array}{ccrcl}
                       \frac{\nu }{2\pi}
                       \rho & 0\leq \rho\leq 2\pi K \\
                 \nu K      &2\pi K \leq \rho \leq 2\pi (K+q)\\
                     \frac{\nu K}{2\pi (P+1-K-q)}(2\pi (P+1) -\rho) & 2\pi (K+q)\leq \rho \leq 2\pi(P+1).
                                             \end{array}
\right.
\end{equation}
  \begin{equation} \label{profileh4exampleIV}
h_4(\rho)
                    =\left\{ \begin{array}{ccrcl}
                       \frac{\beta }{2\pi}
                       \rho & 0\leq \rho\leq 2\pi k\\
               \beta K      &2\pi K \leq \rho \leq 2\pi (K+q) \\
                      \frac{\beta K}{2\pi (P+1-K-q)}(2\pi (P+1) -\rho) & 2\pi (K+q)\leq \rho \leq 2\pi(P+1).
                                             \end{array}
\right.
\end{equation}
The holographic central charge is given by \eqref{centralhol}, that after some algebra yields
\begin{equation}
c_{hol}= 2\beta\nu K^2 (P+2q+1)= 
\left\{ \begin{array}{ccrcl}
                      2\beta\nu K^2 P & \text{if} ~ P>>1 \\
             4\nu  \beta K^2 q      & \text{if}~ q>>1\\
                     2\beta \nu K^2(P+2q+1)& \text{if} ~ K>>1.
                                             \end{array}
\right.
\end{equation}
The comparison with \eqref{centralchargeexampleIV} shows that this is a very stringent check of our proposal.
\\
Finally, the reader can check, using \eqref{numberofbranes}, that the numbers of flavour D8 and D4 branes coincide with the numbers $F_1, F_2$ and $\tilde{F}_1, \tilde{F}_2$ quoted above -- see \eqref{baba}. The same happens with the gauge groups and the numbers of D2 and D6 branes in the associated brane set-up calculated using \eqref{numberofcolour}, and comparing with \eqref{gls}),\eqref{tgls}.\\

Let us now study a qualitatively different example. It will raise a puzzle with an instructive resolution.
\subsection{Example IV: a puzzle and its resolution}
\begin{figure}[h!]
    \centering
    {{\includegraphics[width=14cm]{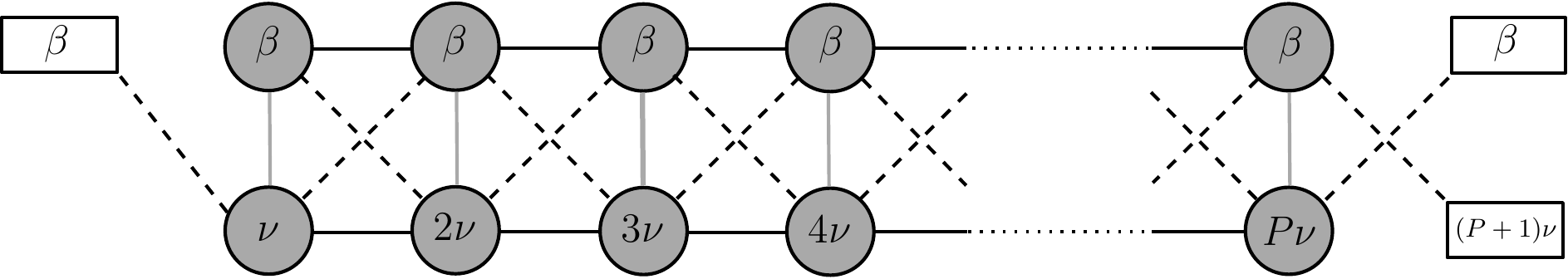} }}%

\caption{ The quiver encoding our fourth example.  The conventions for the fields running along the different lines are the same as those in section \ref{CFTsect}.}
\label{figuraxx}

\end{figure}

Qualitatively, the QFTs discussed above share the fact that the lower row gauge groups 'mirror' the behaviour of the upper row ones. The groups  both grow, stabilise and decrease at the same points. It is interesting to consider an example for which this is not the case. Let us consider the quiver in figure \ref{figuraxx}.

We can easily calculate the number of $(0,4)$ hypermultiplets, vector multiplets and the central charge,
\begin{eqnarray}
& & n_{vec}=P(\beta^2-1) +\sum_{j=1}^{P}\Bigl(j^2 \nu^2-1\Bigr),\;\;\; n_{hyp}= \sum_{j=1}^P j \beta\nu + \sum_{j=1}^{P-1}j(j+1)\nu^2  +\beta^2(P-1),\nn\\
& & c= 3 P^2 (\beta\nu -\nu^2) +(12 +3\beta \nu -3\nu^2)  P- 6\beta^2.\label{feita}
\end{eqnarray}
We can anticipate troubles with the holographic description. Indeed, if we were to take $\nu>\beta$ and large $P$, we could get a negative central charge.
\\
Let us write the functions $h_4, h_8$  describing holographically the IR dynamics of this quiver (as usual $u=\frac{b_0}{2\pi}\rho$),
\begin{equation} \label{profileh8example5}
h_8(\rho)
                    =\left\{ \begin{array}{ccrcl}
                       \frac{\nu }{2\pi}
                       \rho & 0\leq \rho\leq 2\pi P \\
                     \frac{\nu P}{2\pi }(2\pi (P+1) -\rho) & 2\pi P\leq \rho \leq 2\pi(P+1).
                                             \end{array}
\right.
\end{equation}
  \begin{equation} \label{profileh4example5}
h_4(\rho)
                    =\left\{ \begin{array}{ccrcl}
                       \frac{\beta }{2\pi}
                       \rho & 0\leq \rho\leq 2\pi \\
               \beta     &2 \pi \leq \rho \leq 2\pi P \\
                      \frac{\beta}{2\pi }(2\pi (P+1) -\rho) & 2\pi P\leq \rho \leq 2\pi(P+1).
                                             \end{array}
\right.
\end{equation}
The holographic central charge is calculated using \eqref{centralhol}. After some algebra this results in,
\begin{equation}
c_{hol}= 3 P^2 \beta\nu(1+\frac{2}{3P}-\frac{1}{3P^2})\sim 3 P^2 \beta\nu.\label{feita2}
\end{equation}
Comparing the expressions for the field theoretical and holographic central charges in \eqref{feita},\eqref{feita2}, we see a mismatch if we keep the leading order in $P,\nu,\beta$ . This raises a puzzle. The resolution to this puzzle is given by \eqref{lio}. The last interval of the functions $h_4, h_8$ in this example is written as
\begin{eqnarray}
& & h_4^{P,P+1}= \frac{\alpha_P}{2\pi}(2\pi(P+1)-\rho),\;\;\;\; \alpha_P=\beta,\nn\\
& & h_8^{P,P+1}= \frac{\mu_P}{2\pi}(2\pi(P+1)-\rho),\;\;\;\; \mu_P=P \nu .\nn
\end{eqnarray}
Using \eqref{lio}, this implies that the CY$_2$ space is of sub-stringy size, for large $P$. This invalidates the supergravity solution which does not include the dynamics of massless states due to strings or branes wrapping the CY$_2$--see the comment below \eqref{lio}.
The way out of this puzzle is to decouple these light states (by making them heavy and hence the supergravity solution valid). To do this, one must scale $\beta\sim \hat{\beta}\times P$. Then, both the field theoretical and the holographic central charges in \eqref{feita},\eqref{feita2} coincide to $c\sim 3\hat{\beta}\nu P^3$.

We close this section here. A more involved example is discussed in appendix \ref{ejemplodificil}.

\section{Conclusions}\label{conclusion}

This paper presents a new entry in the mapping between SCFTs and AdS-supergravity backgrounds, for the particular case of  two-dimensional small ${\cal N}=(0,4)$  SCFTs and backgrounds with AdS$_3\times$S$^2$ factors. The most general solutions of this type that support an SU(2)-structure on the internal space were recently classified in  \cite{Lozano:2019emq}. \\
We have constructed new solutions of the type AdS$_3\times$S$^2\times$CY$_2$, belonging to class I in the classification in  \cite{Lozano:2019emq}, with compact CY$_2$, whose defining functions are piecewise continuous. We elaborated on their regime of validity and on various general aspects of their mapping with SCFTs. 
In particular, we matched the background isometries and the global symmetries (both space-time and flavour) of the SCFTs. We computed Page charges and put them in correspondence with the putative colour and flavour branes in the  Hanany-Witten set-ups associated to our SCFTs. The CFTs are defined as the IR limit of UV well-behaved long  quivers with $(0,4)$ SUSY, that generalise 2-d (0,4) quivers previously discussed in the literature --see   \cite{Gadde:2015tra,Hanany:2018hlz}. 
Our $(0,4)$ quivers consist of two families of $(4,4)$ quivers coupled by $(0,4)$ and $(0,2)$ matter fields. The $(4,4)$ quivers are associated to D2-NS5-D4 and D6-NS5-D8 brane systems, the latter wrapped on the CY$_2$, which by themselves do not give rise to 2d CFTs in the IR. Our work shows that the coupling between the two families of quivers through matter fields that reduce the supersymmetry to $(0,4)$ renders a 2d CFT in the IR, which admits an AdS$_3$ dual.
After presenting our proposed duality we discussed a
number of examples of increasing complexity that together constitute a stringent test of our proposal. These examples exhibit perfect agreement between the holographic and field theoretical central charges (in the regime where both descriptions are valid), gauge-anomaly cancellation and matching between  isometries  and 'flavour' symmetries on both sides of the duality.
\\
It is clear that this paper just scratches the surface of a rich line of work. In the forthcoming paper \cite{LMNR3} we will apply the developments  in this paper to (among other things) construct a symmetric solution that can be thought of as a completion of the background obtained via non-Abelian T-duality on AdS$_3\times$S$^3\times$CY$_2$. 
Indeed, non-Abelian T-duality has been one of the inspirations of the exhaustive classification presented in \cite{Lozano:2019emq}, and further discussed in this work. This classification provides one more example that shows the huge impact of non-Abelian T-duality as a solution generating technique in supergravity 
--see for example \cite{Lozano:2012au}-\cite{Lozano:2015cra}. One can speculate that an approach similar to the one in \cite{Lozano:2019emq} can be used to classify generic backgrounds in different dimensions and with different amounts of SUSY from particular solutions generated through this technique.

More related to the present paper a number of interesting problems can be tackled. For example, operators of spin two have been studied in correspondence with certain fluctuations of the background metric
\cite{Chen:2019ydk}, \cite{Passias:2016fkm}. It would be interesting to study the analog operators in our CFTs. Similarly, long operators like those in
\cite{Bergman:2018hin} should exist in our CFTs and their associated backgrounds. An obvious open problem is to discuss the CFTs dual to the solutions terminated by the two types of boundary conditions discussed in section \ref{geometria}, not tackled in this paper. 
 In the same vein, it would be interesting to explore the CFT duals of the solutions referred as class II in \cite{Lozano:2019emq}, where the CY$_2$ is replaced by a 4-d Kahler manifold.
\
It would be nice to explore other tests and (more interestingly) find predictions of our proposed duality. The richness of the 2-d SCFTs suggests that stringy tests and mappings along the lines of \cite{Eberhardt:2017pty}-\cite{Dei:2019osr}
 should be possible. We hope to report on these projects soon.

\section*{Acknowledgements}  We would like to thank Giuseppe Dibitetto,  Gaston Giribet, S. Prem Kumar, Daniel Thompson, Alessandro Tomasiello and Stefan Vandoren for very useful discussions.
\\YL and AR are partially supported by the Spanish government grant PGC2018-096894-B-100 and by the Principado de Asturias through the grant FC-GRUPIN-IDI/2018/000174. NTM is funded by the Italian Ministry of Education, Universities and Research under the Prin project ``Non Perturbative Aspects of Gauge Theories and Strings'' (2015MP2CX4) and INFN. CN is Wolfson Fellow of the Royal Society. AR is supported by CONACyT-Mexico. We would like to acknowledge the Mainz Institute for Theoretical Physics (MITP) of the DFG Cluster of Excellence PRISMA$^{+}$ (Project ID 39083149) for its hospitality and partial support during the development of this work. YL and AR would also like to thank the Theory Unit at CERN for its hospitality and partial support during the completion of this work.

\appendix
\section{Appendix: Continuity of the NS sector of our solutions}\label{continuityconditions}
In this section we study the conditions imposed by the continuity of the NS-sector, on the constants $(a_k,b_k,\alpha_k,\beta_k, \mu_k,\nu_k)$ defining our solutions in section \ref{soluciones}.
In particular, we consider solutions that in the interval $[2\pi k, 2\pi(k+1)]$ are given by,
\begin{equation}
\hat{h}_4^{(k)}=\Upsilon \left( \alpha_k +\frac{\beta_k}{2\pi}(\rho-2\pi k)\right),\;\;\;\; h_8^{(k)}= \mu_k +\frac{\nu_k}{2\pi}(\rho-2\pi k),\;\;\;u^{(k)}= a_k +\frac{b_k}{2\pi}(\rho-2\pi k).\label{estacc}
\end{equation}
Below, we quote the value of each component of the metric, $e^{-4\Phi}$ and $B_2$-field when calculated at the point $\rho=2\pi(k+1)$ in terms of the general decomposition
\beq
ds^2= e^{2A}ds^2(\text{AdS}_3)+e^{2C}ds^2(\text{S}^2)+ e^{2D}ds^2(\text{CY}_2)+ e^{-2A}d\rho^2,~~~B= B_0 \text{vol}(\text{S}^2).
\eeq
 If using the solution in \eqref{estacc} we denote them with a superscript $-$. Then, we calculate the NS quantities at the same point $\rho=2\pi(k+1)$, but using the solution in the next interval (with $\alpha_k\to \alpha_{k+1}$, etc), we denote this with a supra-index $+$. Imposing the continuity of each element of the metric and other NS fields, we find conditions for the numbers $(a_k,b_k,\alpha_k,\beta_k, \mu_k,\nu_k)$.
\\
In more detail, we find,
\begin{eqnarray}
& & e^{2A^-}=  \frac{(a_k+b_k)}{\sqrt{\Upsilon(\alpha_k+\beta_k) (\mu_k+\nu_k)}}  ,\;\; e^{2A^+}=\frac{a_{k+1}}{\sqrt{\Upsilon\alpha_{k+1}\mu_{k+1}}}.\\
& & e^{2D^-}= \sqrt{\frac{\Upsilon (\alpha_k+\beta_k)}{(\mu_k+\nu_k)}},\;\;\;\;\; e^{2D^+}= \sqrt{\frac{\Upsilon \alpha_{k+1}}{\mu_{k+1}}}.\nn\\
& & e^{2C^-}= 4\pi^2\frac{(a_k+b_k) \sqrt{\Upsilon(\alpha_k+\beta_k) (\mu_k+\nu_k) }}{b_k^2 +16\pi^2\Upsilon (\alpha_k+\beta_k) (\mu_k+\nu_k)},\;\;\;
e^{2C^+}= 4\pi^2\frac{a_{k+1} \sqrt{\Upsilon\alpha_{k+1} \mu_{k+1} }}{b_{k+1}^2 +16\pi^2\Upsilon \alpha_{k+1} \mu_{k+1}}.\nn\\
& & e^{-4\Phi^-}=\frac{(\mu_k+\nu_k)^3 \left[b_k^2 +16\pi^2\Upsilon  (\alpha_k+\beta_k) (\mu_k+\nu_k)      \right]^2}{256 \pi^4\Upsilon (\alpha_k+\beta_k) (a_k+b_k)^2 },\nn\\
& & e^{-4\Phi^+}=\frac{(\mu_{k+1})^3 \left[b_{k+1}^2 +16\pi^2\Upsilon  \alpha_{k+1}\mu_{k+1}      \right]^2}{256 \pi^4\Upsilon \alpha_{k+1} (a_{k+1})^2 }.\nn\\
& &\frac{ B_0^-}{\pi}=  \frac{a_k b_k - 16\pi^2\Upsilon (\alpha_k+\beta_k) (\mu_k+\nu_k)}{b_k^2+ 16\pi^2\Upsilon (\alpha_k+\beta_k) (\mu_k+\nu_k)},\;\;\;
\frac{B_0^+}{\pi}=  \frac{a_{k+1} b_{k+1}- b_{k+1}^2 - 16\pi^2\Upsilon \alpha_{k+1} \mu_{k+1}}{b_{k+1}^2+ 16\pi^2\Upsilon \alpha_{k+1} \mu_{k+1}}.\nn
\end{eqnarray}
Continuity across $\rho=2\pi(k+1)$ imposes the matching of the analog quantities above. One possible solution is,
\begin{equation}
a_{k+1}= a_k+b_k,\;\;\; b_k=b_{k+1}=b_0,\;\;\; \alpha_{k+1}=\alpha_k+\beta_k,\;\;\; \mu_{k+1}=\mu_k+\nu_k.\label{zaxa}
\end{equation}
These are precisely the same conditions that result from imposing the continuity of  $\hat{h}_4, h_8,u$ across each interval. Notice that \eqref{zaxa} is equivalent to \eqref{conticond}.

\section{A general analysis of Bianchi identities and  counting branes in our Hanany-Witten set-ups}\label{apendice2}
In this appendix we study the charges of D2 and D6  branes induced on D8 and D4 flavour branes. We finish by presenting expressions
to calculate the total number of D8, D6, D4 and D2 branes in a generic Hanany-Witten set-up.
\\
As in the main body of the paper, we denote by $f_p$ the magnetic part of the form $F_p$ and with
$\hat{f}_p$ the magnetic part of the Page field strength $\hat{F}_p= F\wedge e^{-B_2}$.
In the presence of $N_4$ D4 and $N_8$ D8 branes on which we switch a gauge field strength $\tilde{f}_2$ and form the combination ${\cal F }_2= B_2+ 2\pi \tilde{f}_2$. The Bianchi identities read,
\begin{eqnarray}
& & dF_0= \frac{N_8}{2\pi} \delta(\rho-\rho_0) d\rho,\label{bianchid8}\\
& &  df_2-H_3 F_0= \frac{N_8}{2\pi} \delta(\rho-\rho_0){\cal F}_2 \wedge d\rho,\nn\\
& & df_4 -H_3\wedge f_2= (2\pi)^3N_4 \delta(\rho-\rho_0)\delta^4(\vec{y}-\vec{y}_0) d\rho\wedge d^4 \vec{y}+\frac{1}{2}\frac{N_8}{2\pi} \delta(\rho-\rho_0){\cal F}_2\wedge {\cal F}_2 \wedge d\rho,\nn\\
& & df_6- H_3\wedge f_4= (2\pi)^3N_4 \delta(\rho-\rho_0)\delta^4(\vec{y}-\vec{y}_0) {\cal F}_2 \wedge d\rho\wedge d^4 \vec{y} +\frac{1}{6}\frac{N_8}{2\pi} \delta(\rho-\rho_0){\cal F}_2\wedge {\cal F}_2 \wedge {\cal F}_2\wedge d\rho.\nn
\end{eqnarray}
The D8 branes are localised in the $\rho$-direction at the point $\rho_0$, as indicated in the first line of \eqref{bianchid8}. The D4 branes are localised at $\rho=\rho_0$ and at a point $\vec{y}_0$ inside the CY$_2$ space (we denote by $d^4\vec{y}=\text{vol}(\text{CY}_2)$  its volume form).
\\
The explicit definition of the Page field strengths (we only quote the magnetic part here) is,
\begin{eqnarray}
& & \hat{f}_2= f_2-B_2 F_0,\;\;\;\; \hat{f}_4= f_4-B_2\wedge f_2 +\frac{1}{2} B_2\wedge B_2 F_0,\;\;\nn\\
& & 
\hat{f}_6= f_6 - B_2\wedge f_4+\frac{1}{2}
B_2\wedge B_2\wedge f_2-\frac{1}{6}B_2\wedge B_2\wedge B_2 F_0.\label{pagest}
\end{eqnarray}
Combining \eqref{bianchid8} with \eqref{pagest}, we find
\begin{equation}
d\hat{f}_2= N_8 \delta(\rho-\rho_0)\tilde{ f}_2 \wedge d\rho.\label{linda}
\end{equation}
In the case in which there is no gauge field switched on in the D8 branes, there is no induced D6-brane charge, as implied by the first line in \eqref{zada}.
Otherwise D6-{\it flavour} charge is induced, as indicated by \eqref{linda}.
A similar analysis shows that,
\begin{equation}
d\hat{f}_4=(2\pi)^3N_4 \delta(\rho-\rho_0)\delta^4(\vec{y}-\vec{y}_0) d\rho\wedge d^4 \vec{y}+{2\pi N_8} \delta(\rho-\rho_0) \tilde{f}_2\wedge \tilde{f}_2\wedge d\rho 
.
 \end{equation}
This indicates that  D4 brane charge might originate from either localised D4 branes, or on localised D8 branes with a gauge field strength $\tilde{ f}_2$ switched on, such that $\tilde{f}_2\wedge \tilde{f}_2\wedge d\rho $ is non-zero. For our background, we have, consistently with \eqref{secondd}
\begin{equation}
d\hat{f}_4 =(2\pi)^3N_4 \delta(\rho-\rho_0)\delta^4(\vec{y}-\vec{y}_0) d\rho\wedge d^4 \vec{y}.\label{linda2}
\end{equation}
The analogous expression for $\hat{f}_6$ is obtained combining the expressions in \eqref{bianchid8}-\eqref{pagest},
\begin{equation}
d\hat{f}_6=(2\pi)^4  N_4 \delta(\rho-\rho_0)\delta^4(\vec{y}-\vec{y}_0) \tilde{f}_2 \wedge d\rho\wedge d^4 \vec{y}+\frac{1}{6}(2\pi)^2 N_8 \delta(\rho-\rho_0) 
\tilde{f}_2\wedge \tilde{f}_2\wedge \tilde{f}_2\wedge d\rho .
\end{equation}
We thus have $d\hat{f}_6=0$, in agreement with \eqref{zada}.
\\
To close this appendix, let us present simple expressions counting the {\it total}
number of D branes in the Hanany-Witten set-ups associated with our gauge theories and holographic backgrounds. These formulas are similar to those derived in
\cite{Nunez:2019gbg}, \cite{Filippas:2019puw} for CFTs in four and six dimensions. They read,
\begin{eqnarray}
& & N_{D8}^{total}= 2\pi \left[h_8'(0) -h_8'(2\pi (P+1))  \right],\;\;\;\;N_{D4}^{total}= 2\pi \left[h_4'(0) -h_4'(2\pi (P+1))  \right],\label{numberflavourstotal}\\
& &N_{D6}^{total}= \frac{1}{2\pi} \int_{0}^{2\pi(P+1) } h_8 d\rho,\;\;\;\;\;\;\;\;  N_{D2}^{total}= \frac{1}{2\pi} \int_{0}^{2\pi(P+1) } h_4 d\rho.\nn
\end{eqnarray}
These can be successfully checked in all the examples in section \ref{ejempl} and in appendix \ref{ejemplodificil}.

\section{A more stringent check of the duality}\label{ejemplodificil}
\begin{figure}[h!]

    \centering

    {{\includegraphics[width=15cm]{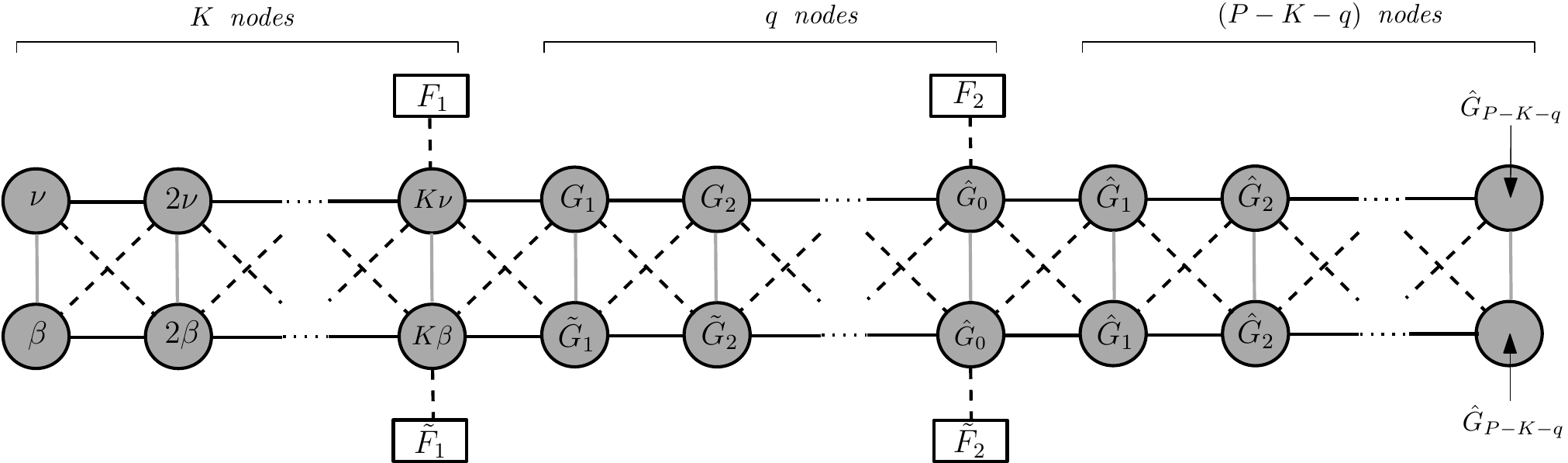} }}%

\caption{A more complicated quiver with $K$ linearly increasing rank nodes in each row, followed by $q$ nodes with $\text{SU}(G_l)$ and $\text{SU}(\tilde{G}_l)$ gauge groups in the top and lower rows, respectively, and ending with $(P-K-q)$ nodes with $\text{SU}(\hat{G}_i)$ gauge groups in both rows.} 

\label{exampleAppen}

\end{figure}

In this Appendix we work out the details of a more complicated, generic and demanding example, shown in figure \ref{exampleAppen}. Extending the examples studied in the body of the paper,  we consider a quiver that starts with linearly increasing nodes. This is followed by $q$-nodes with $\text{SU}(G_l)$, $\text{SU}(\tilde{G}_l)$ gauge groups in the top and lower row respectively, where 
\begin{eqnarray}
\label{Gs5}
G_l=\frac{\hat{G}_0}{q}l + \nu K \left(1-\frac{l}{q}\right),\;\;\;
\tilde{G}_l=\frac{\hat{G}_0}{q}l + \beta K \left(1-\frac{l}{q}\right), \;\;\;\;\;\;l=1,...,q
\end{eqnarray}
Following them, there are $(P-K-q)\;\; \text{SU}(\hat{G}_l)$ gauge groups with ranks
\begin{eqnarray}\label{hatG}
\hat{G}_i&=&\frac{\hat{G}_0}{P-K-q+1}(P-K-q+1-i),\;\;\;\;\;\;i=1,...,(P-K-q)
\end{eqnarray}
in both rows. 

As in the examples studied in the main body of the paper, the gauge anomaly vanishes in the linearly increasing rows. 
Following the logic of section \ref{CFTsect}, for the $\text{SU}(K\nu)$ node we have $Q=F_1+\tilde{G}_1+(K-1)\beta$ and $R=K\beta$.
A vanishing gauge anomaly for the $\text{SU}(K\nu)$ node, see \eqref{noanomaly}, forces
\begin{equation}
F_1= K\beta+\beta-\tilde{G}_1=\frac{\beta}{q}(K+q)-\frac{\hat{G}_0}{q}.
\end{equation}
Similarly, for the $\text{SU}(K\beta)$ node the condition is  
$\tilde{F}_1= K\nu+\nu-G_1=\frac{\nu}{q}(K+q)-\frac{\hat{G}_0}{q}$.

For the next gauge group, $\text{SU}(G_1)$, we have the contibutions $R=\tilde{G}_1$ and $Q=K\beta+\tilde{G}_2$. The gauge anomaly then implies $2\tilde{G}_1-\tilde{G}_2-K\beta=0$, which is true in virtue of \eqref{Gs5}. For all $\text{SU}(G_l)$ and $\text{SU}(\tilde{G}_l)$ gauge groups we also have a vanishing gauge anomaly.
In the $\text{SU}(\hat{G}_0)$ gauge group - $q$-steps forward in the top row- the contributions are $R=\hat{G}_0$ and $Q=F_2+\tilde{G}_{q-1}+\hat{G}_1$, where  \eqref{noanomaly} is satisfied whenever
\begin{eqnarray}
F_2=2\hat{G}_0-\hat{G}_1 -\tilde{G}_{q-1}=\hat{G}_0(\frac{1}{q} +\frac{1}{P+1-K-q}) -\frac{\beta K}{q}.
\end{eqnarray}
The same is true for the $\text{SU}(\hat{G}_0)$ lower gauge group, in this case $\tilde{F}_2=2\hat{G}_0-\hat{G}_1 -G_{q-1}=\hat{G}_0(\frac{1}{q} +\frac{1}{P+1-K-q}) -\frac{\nu K}{q}$.
Considering  \eqref{hatG}, the gauge anomaly vanishes similarly for the rest of the gauge groups.
\\
To calculate the central charge we compute the number of $(0,4)$ hypers and vectors

\begin{eqnarray}
n_{vec}&=&\sum_{j=1}^K \left(j^2 (\nu^2+\beta^2)-2 \right)+\sum_{j=1}^q(G_j^2+\tilde{G}_j^2-2)+\sum_{j=1}^{P-K-q}2(\hat{G}_j^2-1),\nonumber\\
n_{hyp}&=&\sum_{j=1}^K j^2\beta\nu + \sum_{j=1}^{K-1}j(j+1)(\nu^2+\beta^2)  + \sum_{j=1}^{q}G_j\tilde{G}_j +\sum_{j=0}^{q-1}(G_jG_{j+1}+\tilde{G}_j\tilde{G}_{j+1})+\nonumber\\
& &+\sum_{j=1}^{P-K-q}\hat{G}_j^2+\sum_{j=0}^{P-K-q-1}2\hat{G}_j\hat{G}_{j+1},
\end{eqnarray}
where for the number of hypers we are considering $G_0=\nu K$, $\tilde{G}_0=\beta K$ and $\hat{G}_0=G_q=\tilde{G}_q$. 
As in the previous examples, we are interested in the case of a long quiver. To leading order the central charge, in the three possible limits,  is 
\begin{eqnarray} \label{centralchargeexampleV}
c&=&6(n_{hyp}-n_{vec})\nonumber\\
                    &=&\left\{ \begin{array}{ccrcl}
                      2 \hat{G}_0^2P +12P+  O(1, 1/P),\;\; &\text{if} ~P>>1,\\
                       ( 2  \nu \beta K+ (\beta +\nu)\hat{G}_0)K q+ O(1, 1/q),\;\;&\text{if} ~q>>1,\\
                       2\beta \nu K^3  + O(1, 1/K),\;\;&\text{if} ~ K>>1.
\end{array}
\right.
\end{eqnarray}
Now, we can compare the result in \eqref{centralchargeexampleV} with the holographic central charge. The $h_8$ and $h_4$-profiles are given by 
\begin{equation} \label{profileh8exampleIV}
h_8(\rho)
                    =\left\{ \begin{array}{ccrcl}
                       \frac{\nu }{2\pi}
                       \rho & 0\leq \rho\leq 2 \pi K \\
                 \nu K   +\frac{\hat{G}_0-\nu K}{2\pi q}(\rho- 2\pi K)   &2\pi K\leq \rho \leq 2\pi(K+q) \\
                     \hat{G}_0-\frac{\hat{G}_0}{2\pi (P-K-q+1)}(\rho-2\pi (K+q) ) & 2\pi (K+q)\leq \rho \leq 2\pi(P+1).
                                             \end{array}
\right.
\end{equation}
  \begin{equation} \label{profileh4exampleIV}
h_4(\rho) =\left\{ \begin{array}{ccrcl}
                       \frac{\beta }{2\pi}
                       \rho & 0\leq \rho\leq 2\pi K \\
                 \beta K   +\frac{\hat{G}_0-\beta K}{2\pi q}(\rho- 2\pi K)   &2\pi K \leq \rho \leq 2\pi(K+q) \\
                     \hat{G}_0-\frac{\hat{G}_0}{2\pi (P-K-q+1)}(\rho-2\pi (K+q) ) & 2\pi (K+q)\leq \rho \leq 2\pi(P+1).
                                             \end{array}
\right.
\end{equation}
The holographic central charge, using \eqref{centralhol}, results into
\begin{eqnarray}
\label{CCExtraAppenExample}
c_{hol}&=& 2\beta\nu K^2(K + q) +(\beta+\nu)K q \hat{G}_0 - 2 \hat{G}_0^2(K-P-1)\nonumber\\
&=& 
\left\{ \begin{array}{ccrcl}
                      2 \hat{G}_0^2 P &\text{if}~P>>1,\\
             (2\nu \beta K + (\beta+\nu)\hat{G}_0 ) Kq &\text{if}~ q>>1 \\
                     2\beta\nu K^3 &\text{if}~K>>.                                             \end{array}
\right.
\end{eqnarray}
We can then easily see that \eqref{CCExtraAppenExample} is in complete agreement with the output of \eqref{centralchargeexampleV}.

\end{document}